\documentclass[11pt,a4paper]{article}
\pdfoutput = 1
\usepackage{jcappub}
\usepackage[utf8]{inputenc}
\usepackage{amsmath,amssymb}
\usepackage{graphicx}
\usepackage{hyperref}
\usepackage{natbib}
\usepackage{mathtools}
\usepackage{subcaption}

\newcommand{\bq}{\textbf{q}}
\newcommand{\bx}{\textbf{x}}
\newcommand{\bk}{\textbf{k}}

\newcommand{\cdg}{\ensuremath{C_{\ell}^{\delta_{g, \rm obs}^{i}, \gamma_{E}^{j}}}}
\newcommand{\ckk}{\ensuremath{C_{\ell}^{\kappa^i\kappa^j}}}
\newcommand{\cgg}{\ensuremath{C_{\ell}^{\gamma_{E}^{i}\gamma_{E}^{j}}}}

\newcommand{\mo}{\mathcal{O}}

\renewcommand{\vec}{\mathbf}

\newcommand{\photoz}{photo-$z$}
\newcommand{\lephare}{\textsc{LePHARE}}
\newcommand{\trudircal}{\textit{TrueDirCal}}
\def\kMpc{\, h \, {\rm Mpc}^{-1}}

\newcommand{\threetimestwo}{$3\times2$-point}
\newcommand{\twotimestwo}{$2\times2$-point}

\author[a,b]{Joseph DeRose,}
\author[b,c]{Noah Weaverdyck,}
\author[b,c,d]{Martin White,}
\author[e,f]{Shi-Fan Chen,}
\author[c]{David Schlegel,}
\author[a]{An\v{z}e Slosar}

\affiliation[a]{Physics Department, Brookhaven National Laboratory, Upton, NY 11973, USA}
\affiliation[b]{Berkeley Center for Cosmological Physics, Department of Physics, UC Berkeley, CA 94720, USA}
\affiliation[c]{Lawrence Berkeley National Laboratory, 1 Cyclotron Road, Berkeley, CA 93720, USA}
\affiliation[d]{Department of Physics, University of California,Berkeley, CA 94720}
\affiliation[e]{Department of Physics, Columbia University, New York, NY 10027, USA}
\affiliation[f]{NASA Hubble Fellowship Program, Einstein Fellow}
\emailAdd{jderose@bnl.gov}

\title{Steeling Weak Lensing Source Galaxy Samples against Systematics using Wide Field Spectroscopy}


\abstract{
We investigate the cosmological constraining power of combined weak galaxy lensing and galaxy clustering probes, i.e.\  $3\times2$-point analyses, assuming flexible models for redshift uncertainty, and Lagrangian perturbation theory and hybrid effective field theory models for galaxy intrinsic alignments, galaxy bias and baryonic physics. In this context, we provide a detailed accounting of the limiting systematics on $3\times2$-point analyses. Our main finding is that in the presence of current levels of uncertainty on baryonic physics, the information content of weak lensing analyses saturates on quasi-linear scales, allowing the use of source galaxy samples that are significantly less dense, e.g.\ with number densities of $5\rm \, arcmin^{-2}$, without sacrificing constraining power, provided that redshift distributions can be calibrated at the $\sigma(\langle z\rangle)=0.005$ level. We show that for sufficiently narrow lens and source redshift distributions, intrinsic alignment contributions can be largely self-calibrated, though sufficient flexibility must be given to the redshift and scale dependence of this signal. The near optimality of such relatively sparse source galaxy samples opens the possibility to directly calibrate the redshift distributions and intrinsic alignment contamination of such a sample using a spectroscopic instrument like DESI, thus mitigating the dominant systematics in weak lensing analyses.
}
\begin{document}
\maketitle
\flushbottom

\section{Motivation}

Galaxy weak lensing is a powerful probe of the low redshift universe, sensitive to the total matter distribution and the expansion history of the universe. New experiments such as the Rubin Observatory \cite{Ivezic2008} and Euclid Telescope \cite{Laureijs2011} were designed in large part to measure galaxy lensing statistics and their cross-correlations with galaxy densities, and have begun observations over the last two years. If these surveys deliver on their promised data quality, the cosmology community will soon be sitting on a treasure trove of lensing data that holds the promise to weigh in on many open questions in cosmology, including cross-checking recent measurements of the redshift evolution of the dark energy equation of state, probing the nature of dark matter and constraining light relics and neutrinos. But before the cosmology community can reap the benefits of these data, several key systematic uncertainties in galaxy lensing analyses must be addressed \cite{Mandelbaum2017,Boruah2023,Zhang2026,Robertson2026}, including redshift calibration, galaxy intrinsic alignments and the impact of galaxy formation on the small scale matter distribution.

The flagship analyses that galaxy lensing surveys have pursued to date are the $3\times2$-point analyses, so called because they combine the three two-point statistics that can be formed between the (projected) galaxy over-density field and galaxy shear field. These three statistics are the auto-power spectrum of galaxy over-densities (galaxy clustering), the cross-power spectrum between galaxy over-density and galaxy shear (galaxy-galaxy lensing) and the auto-power spectrum of galaxy shear (cosmic shear). The galaxies used to construct the shear field entering into these measurements are often called source galaxies, and those used to construct the over-density field are referred to as lens galaxies. Furthermore, it is standard practice to sub-select source and lens galaxies as a function of redshift to more finely probe the redshift evolution of the physics being studied with \threetimestwo\ analyses.  As we will see later, the ability to form narrow and well-defined redshift distributions will offer several advantages.

One of the major hurdles that Stage IV \threetimestwo\ analyses will face is the lack of representative spectroscopic data for redshift distribution calibration. The redshift distributions of lens and source galaxies are a key input for models of galaxy clustering and weak lensing. Current generation surveys use a combination of spectroscopic data \cite{Hartley2020,Masters2019,McCullough2024,Lange2026,Blanco2026}, medium-band photometric data \cite{PAUS,Shuntov_2025}, and cross-correlation redshifts \cite{Gatti2018,Myles2021,Rau2023,dAssignies2025,Choppin2026,Ruggeri2026} to constrain weak lensing redshift distributions. The recent Hyper Suprime-Cam Year 3 analyses \cite{Dalal2023,Li2023}, using data with comparable depths to the expected first year of Rubin data, have identified an inability to sufficiently calibrate their redshift distributions using such an approach \cite{Rau2023} as a major source of error in their analysis \cite{Dalal2023}. 

Stage III surveys' reliance on less accurate redshift calibration techniques such as medium band photometry and clustering redshifts is driven by known failure modes when applying existing, heterogeneous catalogs of spectroscopic data to the calibration weak lensing sources. Current catalogs of spectroscopic redshifts 
lack representation of faint objects, where it is difficult to achieve secure redshifts,
and challenges can arise even for brighter samples because of the difficulties in modeling the complex selection function of existing spectroscopic datasets \cite{Hartley2020, crafford2026diagnosingeffectsspectroscopictraining}. Significant efforts have been undertaken to characterize the color-redshift relation using spectroscopic observations \cite{Masters2019,McCullough2024}, and while these efforts have provided large advances in the state of the field, whether this technique suffices for Euclid \cite{Masters2019} or Rubin Year 1 depths \cite{Robertson2026} still hinges on the assumption that the mean redshift of a sample of galaxies at fixed color either doesn't evolve with magnitude or evolves in a well-defined and precisely measured manner.

The most reliable way of calibrating the redshift distribution for a galaxy population is simply to take spectroscopic redshift measurements of objects sampled directly from it. Such a dedicated, direct calibration program is now much more feasible, given technological advances in large-scale spectroscopy, and can reveal unexpected features in redshift distributions that are not well captured by the limited uncertainty characterization of photo-$z$ approaches \cite{trudircal}. The challenge of this approach is that one needs to achieve very high levels of spectroscopic completeness, meaning that an unbiased set of redshifts must be obtained for a representative selection of galaxies in the sample. Given the requirement of $\Delta z_{\rm mean} < 0.001(1+z)$ in the Rubin Observatory Legacy Survey of Space and Time (LSST) Dark Energy Science Collaboration (DESC) Science Requirements Document (SRD) \cite{Mandelbaum2017}, even levels of spectroscopic incompleteness comparable to those seen for significantly brighter samples than the Rubin Gold sample would cause this requirement to fail. 

Recent measurements with DESI suggest that achieving very high redshift completeness for the $i$-band-magnitude-limited Rubin samples will be out of reach even with very long exposure times.
This is driven by the extreme faintness of the bulk of Gold sample galaxies, the fact that some galaxies do not display prominent emission lines, or that there are no prominent spectroscopic features that fall within the wavelength range of the DESI spectrographs for galaxies in the so-called ``redshift desert'', $1.6<z<2.1$, where [O{\sc ii}] has redshifted out and before Ly$\alpha$ has redshifted into the DESI wavelength range. 
Obtaining redshifts for galaxies in this redshift range requires spectrographs with sensitivity further in the UV or infrared \cite{Steidel04}, or achieving spectroscopy at (approximately) five times higher signal-to-noise to fit the absorption features including the [Mg{\sc ii}] and the [Fe{\sc ii}] complex, requiring a factor of approximately 25 times as much effective observing time.

While it may be possible to \textit{post-facto} remove the portion of Gold sample galaxies for which we have no spectroscopic calibration during the weak lensing analysis, doing so would likely require removing a significant fraction of galaxies from this sample. A much more efficient approach in terms of spectroscopic resources is to select a parent sample for ``true" direct calibration\footnote{The term \textit{direct calibration} has been used in several ways in the literature.  For example, it can refer to reweighting a pre-existing, incomplete and non-representative spectroscopic sample to match the photometric properties of the photometric galaxy sample. Here we \textit{actually} mean ``direct'', in the sense of obtaining spectroscopic observations from the target sample directly, hence ``true" direct calibration, or \trudircal{}.}, wherein nearly all of the galaxies can be redshifted by DESI or a similar instrument. This approach is proposed and advocated for in Ref.~\cite{trudircal} under the moniker \trudircal{}, where it is tested on lens samples from the Dark Energy Survey. In this work, we investigate applying this to source galaxy samples from upcoming surveys, showing that such a sample could be competitive with much denser galaxy samples in terms of cosmological constraining power, and that a sample with sufficient density can likely be calibrated with DESI. 

Calibrating redshift distributions of faint sources via so-called clustering redshifts is also subject to systematic errors, such as source galaxy bias evolution and magnification, which are difficult to control at the required level.
Progress is being made on clustering redshift techniques, particularly for precisely calibrating the mean redshifts of source samples, but the ability to calibrate widths of redshift distributions and outlier contributions at the same level of precision still lags significantly behind what is required, even in relatively optimistic simulation tests \cite{dAssignies2025}. As we will see later, precise calibration of the widths of source redshift distributions and, to a lesser extent, outlier fractions will be necessary to achieve optimal \threetimestwo\ constraints with LSST.

Even if deep spectroscopy can be obtained, or redshift distributions can be sufficiently calibrated via clustering redshifts, the width of the redshift bins used for weak lensing analyses will dictate the amount of intrinsic alignment contamination in lensing measurements. Intrinsic alignments are an effect whereby unlensed, ``intrinsic'' galaxy shapes are aligned with each other and with the local matter distribution \cite{Lamman2024,Siegel2026a}. At a given redshift, the linear intrinsic alignment contribution is perfectly degenerate with a change in the amplitude of the lensing signal ($S_8$).
Fortunately IAs are known to be relatively small \cite{Johnston2019,Samuroff2023,Siegel2026a}, and in fact for Stage I and Stage II surveys could be safely neglected. For Stage III surveys even relatively crude models were sufficient and marginalization over IA uncertainties changed the constraining power by 5-40\% depending upon the rigidity of the assumed model \cite{Wright2025,Abbot2026,Chen2024b}. However as we enter the Stage IV era a proper modeling or mitigation of IAs becomes critical, since even relatively small effects can become statistically important \cite{Blazek2019,Samuroff2024,Blot2025}.  One way that past galaxy lensing analyses broke the $S_8$-intrinsic alignment degeneracy was by assuming an under-parameterized form for the redshift evolution of intrinsic alignments, often a power-law, which can then be constrained using different redshift bin combinations. This simplicity is dangerous. If the redshift evolution in the Universe does not conform to the assumptions in these models, then this degeneracy will be incorrectly broken leading to biases in cosmological parameters. 

A more robust intrinsic alignment mitigation strategy is to reduce the width of source galaxy bins. The intrinsic alignment contribution to galaxy-galaxy lensing signals is directly proportional to the overlap between source and lens redshift distributions, thus intrinsic alignment contamination can be reduced in a model independent manner by reducing this overlap \cite{Samuroff2024,Chen2024b}. On the other hand, the intrinsic alignment contributions to cosmic shear measurements are very difficult to suppress in a model independent way. An uncertainty on the linear alignment amplitude of $\sigma(A_{\rm IA})=1$ leads to $\sim1-2\%$ uncertainties for the least contaminated, highest redshift auto-spectra, increasing to significantly larger uncertainties for lower redshift and cross-bin cosmic shear spectra. Thus to make use of the cosmic shear spectra, constraints on alignments with $\sigma(A_{\rm IA})<1$ for the samples of interest will be required. Recent analyses have attempted to alleviate this requirement by using existing spectroscopic measurements of IAs to tailor samples of galaxies for which the intrinsic alignment contamination is expected to be small, e.g.\ by selecting blue samples of galaxies \cite{McCullough2024b,Bigwood2026}, assuming that the IA signal does not evolve unexpectedly in the regions of photometric space not probed by spectroscopy. An alternative avenue for obtaining these constraints is through \twotimestwo\ analyses. If one can make the source galaxy redshift distributions compact, then simpler IA redshift evolution models can be used, which can then be directly constrained in a \twotimestwo\ analysis, thereby allowing significant amounts of information to be extracted from cosmic shear. 

A further component is required to make full use of the shear auto-spectrum: a model for the matter power spectrum. The forecasts in the DESC SRD assume that the matter power spectrum model is accurate to $\ell=3000$, which translates to $k\sim2\, \kMpc$ at $z\sim0.5$, where the lensing contributions in Rubin analyses will peak. A major complicating factor is that, in addition to the gravitational non-linearities that impact these scales, there are significant contributions from non-gravitational physics, often referred to as baryonic feedback. Recent measurements of the gas profiles of galaxies making use of DESI combined with high resolution CMB measurements \cite{Hadzhiyska2025,Guachalla2025}, which the current generation of hydrodynamical simulations were not tuned to fit, appear to challenge the prevailing picture that these feedback effects are weak.  The preferred gas profiles are consistent with simulations that predict baryonic feedback effects of $\sim 20\%$ on the matter power spectrum at $k\sim2\,\kMpc$.  (The thermal pressure profiles of groups and clusters predicted by current-generation simulations and models also appear to be in conflict with CMB measurements \cite{Efstathiou25,Popik26,Liu26}.)  This renders the interpretation of weak lensing measurements probing these scales extremely challenging. 
Furthermore, the observations against which these simulations can be calibrated have proven difficult to measure (e.g.\ \cite{Akino2022,Kluge2024,Eckert2026}) and reliably interpret (e.g.\ \cite{Moser2021}). Efforts to use these observations to mitigate the uncertainties of baryonic feedback on weak lensing will be highly sensitive to mis-modeling the impact of baryonic physics on the matter power spectrum \cite{Robertson2026}. For recent summaries, with additional references to the original literature, see refs~\cite{Bartlett2026,Bigwood2026,Siegel2026b}.

\begin{figure}
    \centering
    \includegraphics[width=\linewidth]{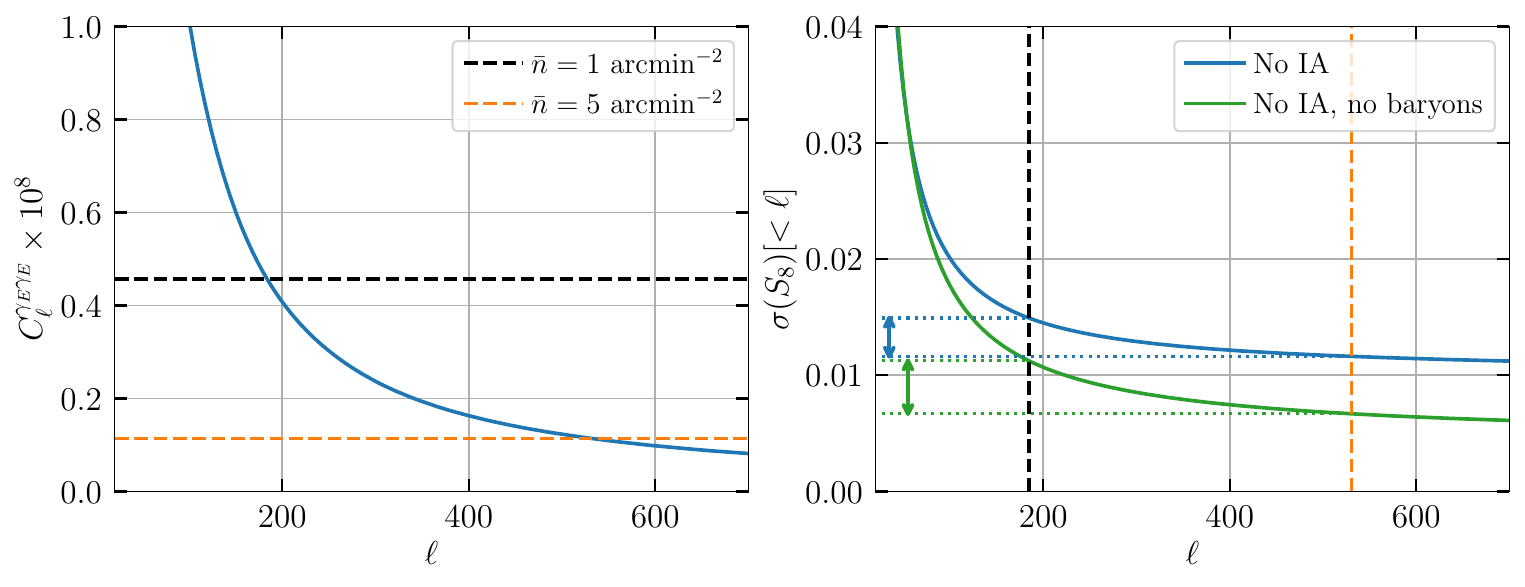}
    \caption{({\it Left}) Cosmic shear power spectrum for the highest redshift Steel sample bin (solid) compared to shape noise contributions per mode for different source number densities (dashed), roughly corresponding to the densities assumed throughout this paper for Steel (black) and Gold (orange) samples. ({\it Right}) Fisher forecasts for $\sigma(S_8)$ as a function of $\ell_{\rm max}$ without shape noise, marginalizing over \photoz\ and shear calibration uncertainties and fixing IA uncertainties to zero (blue) or IA and baryon uncertainties to zero (green). The vertical lines indicate where the lensing signal is greater than the shape noise for the two source densities plotted on the left. When baryons are marginalized over, the vast majority of the constraining power on $\sigma_8$ comes from scales where the cosmic shear signal is not shape noise dominated, even for relatively low source number densities. This is seen by the relatively small improvement in $\sigma(S_8)$ between the scales that are shape noise dominated for the two different densities, as indicated by the blue arrow. When baryons are fixed, more information can be extracted from high $\ell$, although even in this case the rapidly decreasing lensing signal at high $\ell$ limits the advantage of very high number densities.}
    \label{fig:baryon_comp}
\end{figure}

As shown in Figure~\ref{fig:baryon_comp}, one of the main consequences of astrophysical uncertainties on small scales is that, once the uncertainty is accounted for, the cosmological constraining power of galaxy lensing analyses shifts to larger scales.  The clustering power on these scales is larger, and thus sample variance limited constraints are achieved at lower galaxy densities than those originally proposed for Stage IV source galaxy samples. This presents an opportunity to re-weight the contributions of statistical and systematic errors in weak lensing analyses. Such an approach has also been suggested by Ref.~\cite{McCullough2024b}, though their focus is primarily on reducing IA contamination and ours will be on redshift calibration.  In particular, here we investigate the possibility of using a sample of galaxies for which secure redshifts can be obtained via a wide--field spectroscopic instrument, e.g.\ DESI, and that can be selected with high purity using broadband fluxes.

As we shall see below, spectroscopically calibrated lens galaxy samples of sufficient density already exist, so the main challenge is the characterization of the source samples.  The main result of this work is that a lower density sample of source galaxies with well-calibrated redshifts can yield cosmological constraints that exceed the constraining power of the Gold sample proposed in the DESC SRD, when assuming uninformative priors on scale-dependent bias and intrinsic alignments, even with significant improvements in redshift uncertainty (although not quite meeting the requirements in the SRD). A forthcoming companion paper \cite{steel_obs} demonstrates the feasibility of efficiently observing, selecting, and characterizing such a sample with DESI. Furthermore, in this work we focus on the impact of various choices on $S_8$ constraining power within $\Lambda$CDM and defer forecasts for extended parameter spaces to a follow up publication.

The outline of the rest of this paper is as follows. In Section~\ref{sec:data} we specify our assumptions about DESI and LSST data. Section~\ref{sec:theory_model} describes the theoretical model that is used in this work, and Section~\ref{sec:framework} describes Fisher forecasting framework for \threetimestwo\ analyses using this model.  Our forecasts are presented in Section~\ref{sec:constraints}.  Section~\ref{sec:feasibility} investigates the feasibility of the Steel sample program outlined in the previous sections. We summarize with an outlook in Section~\ref{sec:outlook}. We relegate significant additional details about our intrinsic alignment modeling, and forecasts for direct intrinsic alignment constraints, including requirements on spectroscopic survey area and number density, to the Appendices.

\section{Data Assumptions}
\label{sec:data}

We now outline the assumptions that we have used to produce forecasts for the \threetimestwo\ analyses in this work. In particular, we assume DESI-like galaxies as lenses and Rubin galaxies as sources. We use flexible parameterizations for galaxy bias, intrinsic alignments and baryonic effects as described below.

\subsection{DESI}
\label{sec:desi}

For the lenses in our forecasts we use the DESI bright galaxy sample (BGS) and luminous red galaxies (LRGs) selected over the full LSST area, i.e.\ 14,300 square degrees, the area assumed for LSST Y10 in Ref.~\cite{Mandelbaum2017}, covering $0.1<z<1.1$. This is enabled by the nearly total coverage of the LSST footprint by the DESI legacy imaging surveys \cite{Dey2019}. For BGS and LRGs, we use two and four photometrically-selected but spectroscopically calibrated redshift bins following \cite{Chen2024b,Sailer25a,Sailer25b,Maus25}.  The properties of these samples are summarized in Table \ref{tab:lens_info}.  In what follows, we will assume there are no uncertainties in the lens redshift distributions as these samples have $99\%$ spectroscopic completeness \cite{Hahn23,Zhou23} and are spectroscopically calibrated over the full DESI footprint, leading to negligible sample variance. 

\begin{table*}[t!]
    \centering
    \begin{tabular}{|c|c|c|c|c|c|c|c|c|}
         \hline
         \hline
         Sample & $z_{\rm eff}$ & $b_{1E}$ & $\alpha_{\mu}$ & $10^{6}\,\textrm{SN}_{\rm 2D}$ & $\textrm{SN}_{\rm 3D}$ & $\bar{n}$ & $\ell_{\rm max,fid}$ & $\ell_{\rm SN}$ \\
         & & & & & [$h^{-3}$Mpc$^3$] & $[\textrm{arcmin}^{-2}]$ & & \\
         \hline
         BGS0 & 0.211 & 0.99 & 2.025& 0.463 & 90   & 0.174 & 134 & 1950 \\
         BGS1 & 0.352 & 1.34 & 2.0  & 0.918 & 430  & 0.088 & 267 & 1050 \\
         LRG0 & 0.470 & 1.72 & 2.43 & 3.89  & 2835 & 0.021 & 400 & 480 \\
         LRG1 & 0.625 & 1.96 & 2.61 & 2.16  & 2600 & 0.038 & 533 & 640 \\
         LRG2 & 0.785 & 2.73 & 2.44 & 2.03  & 3350 & 0.041 & 667 & 680 \\
         LRG3 & 0.914 & 2.47 & 1.95 & 2.24  & 5295 & 0.038 & 767 & 510 \\
         \hline 
    \end{tabular}
    \caption{Summary of quantities pertaining to the DESI lens samples used in this work.
    $z_{\rm eff}$ is the effective redshift, $b_{1E}$ the Eulerian linear bias, $\alpha_{\mu}$ the lens magnification coefficient given by Eq.~\ref{eq:magnification_coeff}, $\textrm{SN}_{\rm 2D}$ the Poisson angular shot noise, $\textrm{SN}_{\rm 3D}$ the best-fit three-dimensional shot noise, allowing for deviations from the Poissonian expectation from \cite{Chen2024b}, $\bar{n}$ the angular number density, and $\ell_{\rm max,fid}$ the $\ell$ value corresponding to $k_{\rm max}=0.4\kMpc$, which is the maximum wavenumber used for clustering and galaxy-galaxy lensing in our forecasts.  The final column, $\ell_{\rm SN}$, lists the harmonic at which the total angular power is twice the Poisson shot-noise value, which is where the clustering and shot-noise are equal. }
    \label{tab:lens_info}
\end{table*}

As shown in Fig.~\ref{fig:nz} these samples have narrow $n(z)$ with negligible tails, which is highly advantageous in controlling systematic errors in the analysis -- a point we shall return to later.  They are also drawn from well-calibrated imaging that can be cross-checked against full 3D clustering from the DESI spectroscopic samples, providing a ``low systematics'' clustering sample.  As we shall see later, the density of these lenses is sufficiently high that their shotnoise will be a subdominant contribution to the error bar in any auto- or cross-spectrum that involves these samples over the whole range of scales that we consider.  This suggests there is little obvious gain in using denser samples.  However, should an enhanced density become beneficial, spectroscopic coverage of the ``luminous galaxy extension'' of the samples we consider will be available with DESI DR3. This increases the density of the LRG samples by around 50\%, while maintaining high spectroscopic purity and coverage.  

While we shall not consider it in this work, we note that the use of DESI galaxies as lenses would enable the angular clustering of the lens sample to be replaced with 3D, redshift-space clustering in future (e.g.\ paralleling what was done in Refs.~\cite{Chen22,Maus25} with CMB lensing).

\subsection{LSST}
\label{sec:lsst}

\begin{figure}
    \centering
    \includegraphics[width=\linewidth]{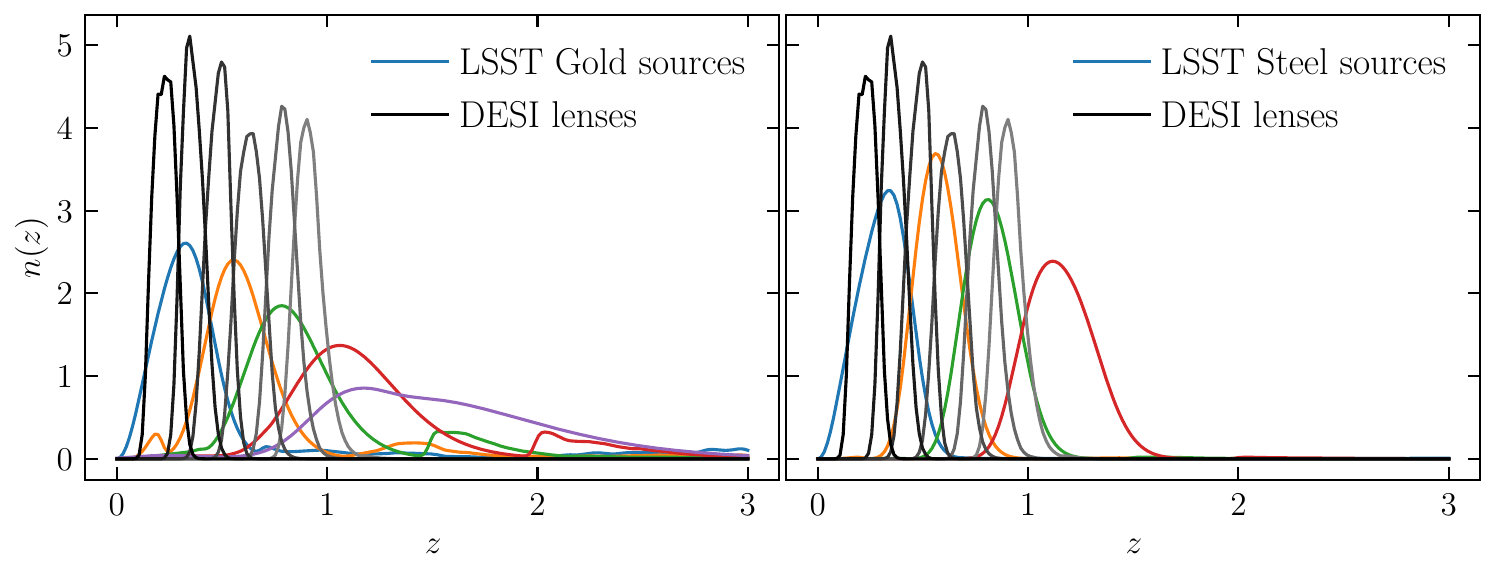}
    \caption{Redshift distributions (arbitrarily normalized) assumed for LSST sources (colored lines) for Gold (left) and Steel (right) samples constructed as described in Section~\ref{sec:lsst}.  The BGS and LRG lens bins (greyscale; \S\ref{sec:desi}) are the same in both panels.  Note that the Steel sample $n(z)$ are more compact, but have negligible support above $z=1.6$, where [O{\sc ii}] redshifts out of the DESI spectrograph wavelength range.
    }
    \label{fig:nz}
\end{figure}

We have assumed an LSST survey area of 14,300 square degrees, i.e.\ $f_{\rm sky}=0.34$, following \cite{Mandelbaum2017}. For our LSST ``Gold'' sample forecasts we assume a number density of 27.7~$\rm arcmin^{-2}$ to $i_{\rm mag}=25.3$, divided into five equal-density redshift bins. These are constructed by convolving a true underlying redshift distribution with a Gaussian error:
\begin{align}
\label{eq:nz_core}
    p^{\gamma^{i}}_{\rm core}(z) &= N^{-1}\int_{z_{\rm min}}^{z_{\rm max}} dz_{\rm p}\ n_{\rm true}(z)\, p(z|z_{p}) \nonumber\\
    &= N^{-1}\int_{z_{\rm min}}^{z_{\rm max}} dz_{\rm p}\ z^{\alpha}\, e^{-(z/z_0)^\beta} \mathcal{N}(z_p,\sigma_{z,i}^2)\, ,
\end{align}
where $z_{\rm min}$ and $z_{\rm max}$ are the edges of the photometric redshift bin, $\sigma_{z,i}$ is the assumed photometric redshift error, and $\alpha=2$, $z_0=0.179$ and $\beta=0.798$ as described in Refs.~\cite{Fang2021,Zhang2026}, and $N^{-1}$ is a normalization constant set such that $p^{\gamma^{i}}_{\rm core}(z)$ integrates to one. The bin edges that result in equal true number densities per bin are $z=\{0, 0.45, 0.69, 0.98, 1.39, 2.98\}$. For the Gold sample, we use the values in Table 1 of ref.~\cite{Zhang2026} for $\sigma_{z,i}$, i.e.\ $\sigma_{z,i}=\{0.065, 0.0825, 0.0975, 0.1175, 0.1875\}$. For the Steel sample we assume the same parametric form, independent of density, but set $\sigma_{z,i}=0.05(1+z)$ motivated by the fact that this sample is significantly brighter than the Y10 Gold sample. Furthermore, we set $z_{\rm max}=1.5$ for the highest redshift Steel sample source bin in order to account for the fact that DESI will not be able to efficiently measure redshifts above $z\approx 1.6$. This roughly aligns with the preliminary $n(z)$s obtained with actual DESI data, shown in our forthcoming companion paper \cite{steel_obs}.

We include an additional outlier component, $p^{\gamma^{i}}_{\rm out}(z)$, which is estimated from the \texttt{cosmoDC2} simulation \cite{Korytov2019} following the procedure outlined in Ref.~\cite{Zhang2026}, yielding a total redshift distribution for each tomographic bin of:
\begin{equation}
\label{eq:f_out}
    p^{\gamma^{i}}(z) = (1 - f_{{\rm out},i})\,p^{\gamma^{i}}_{\rm core}(z) + f_{{\rm out},i}\,p^{\gamma^{i}}_{\rm out}(z)\, ,
\end{equation}
\noindent where $f_{{\rm out},i}$ is the assumed outlier fraction.  We take $f_{{\rm out},i}=0.15$ for the Gold sample, and $0.01$ for the Steel sample in our fiducial model, although we will leave this as a free parameter for both samples as discussed in Section~\ref{sec:obs_sys}. This outlier fraction is distinct from the redshift success rate, $f_{\rm spec}$, discussed in Section~\ref{sec:feasibility}, and there is reason to believe that $f_{\rm out}$ will be significantly less than $1-f_{\rm spec}$. Regardless, we will see below that we are not particularly sensitive to the mean outlier fraction, although we are sensitive to its uncertainty. For all source samples, we assume $\sigma_{e}=0.26$. Figure \ref{fig:nz} shows the assumed LSST Gold and Steel sample redshift distributions along with the DESI redshift distributions described above.

\section{Theoretical Model}
\label{sec:theory_model}

In order to make quantitative arguments about how weak lensing source galaxy samples should be optimized, we must specify a model for the \threetimestwo\ data vector. In this work, we will focus on modeling the auto- and cross-power spectra of two types of fields: the observed, projected galaxy density, $\delta_{g, \rm obs}^{i}$, and the $E$-mode of the galaxy ellipticity field, $\gamma_{E}^{j}$, where the superscripts on these fields index over different redshift bins. From these we can form three distinct types of angular power spectra, namely galaxy clustering, $C^{\delta_{g, \rm obs}^{i},\delta_{g, \rm obs}^{j}}_{\ell} $, galaxy-galaxy lensing, $C^{\delta_{g, \rm obs}^{i}, \gamma_{E}^{j}}_{\ell}$, and cosmic shear $C^{\gamma_{E}^{i}, \gamma_{E}^{j}}_{\ell}$. We now specify the theoretical contributions to these signals.

The observed galaxy density field can be expressed as 
\begin{equation}
    \delta_{g,\rm obs} = \delta_{g} + \delta_{g,\mu} \\
\end{equation}
\noindent where $\delta_{g}$ is the intrinsic projected galaxy overdensity, $\delta_{g,\mu}$ is the lens magnification contribution to the observed galaxy density, and we neglect the redshift-space distortion contribution given that we are projecting over large kernels. There are some limits in which the RSD term becomes important \cite{Fang2020}, but in those cases neglecting this contribution will lead to biases rather than changes in the constraining power of the analysis which is the main interest of this work. 

The other field of interest can be expressed as
\begin{equation}
    \gamma_{E} = \kappa + \gamma_{E,I}\, ,
\end{equation}
\noindent where $\kappa$ is the true lensing contribution (convergence) to the $E$-mode of the observed galaxy ellipticity and $\gamma_{E,I}$ is the intrinsic alignment contribution. We neglect higher-order lensing contributions, although these are expected to become important at the precision of the LSST data \cite{Krause2021}. These are mostly deterministic calculations with no free parameters, other than source galaxy magnification terms, thus we expect their inclusion to negligibly impact the constraining power of these analyses.

\subsection{Angular Power Spectra}

Throughout this work, we will use the Limber approximation to compute angular power spectra between fields $a$ and $b$:
\begin{equation}
\label{eq:limber}
    C^{ab}_{\ell} = \int d\chi\ \frac{W^{a}(\chi)W^{b}(\chi)}{\chi^2} 
    P_{ab}\left( k_{\perp}=\frac{\ell + \frac{1}{2}}{\chi}, k_{\parallel}=0 ; z(\chi)\right) + \mathcal{O}(\ell^{-2}) ,
\end{equation}
\noindent where $W^a$ and $W^b$ are projection kernels specific to the fields $a$ and $b$ respectively and are listed below, and $P_{ab}(k)$ is the three-dimensional (cross-)power spectrum of these fields. Given the field level descriptions of $\delta_{g,\rm obs}$ and $\gamma_{E}$ above, it is clear that the auto- and cross-spectra of these fields receive the following contributions:

\begin{align}
C^{\delta_{g, \rm obs}^{i},\delta_{g, \rm obs}^{j}}_{\ell} &= C^{\delta_{g}^{i},\delta_{g}^{j}}_{\ell} + C^{\delta_{g}^{i},\delta_{g, \mu}^{j}}_{\ell} + C^{\delta_{g}^{j},\delta_{g, \mu}^{i}}_{\ell} + C^{\delta_{g, \mu}^{i},\delta_{g, \mu}^{j}}_{\ell} \nonumber \\
C^{\delta_{g, \rm obs}^{i}, \gamma_{E}^{j}} &= C^{\delta_{g}^{i},\kappa^{j}}_{\ell} + C^{\delta_{g}^{i},\gamma_{E,I}^{j}}_{\ell} + C^{\delta_{g, \mu}^{i},\kappa^{j}}_{\ell} +  C^{\delta_{g, \mu}^{i},\gamma_{E,I}^{j}}_{\ell} \\
C^{\gamma_{E}^{i}, \gamma_{E}^{j}} &= C^{\kappa^{j},\kappa^{j}}_{\ell} + C^{\kappa^{i},\gamma_{E,I}^{j}}_{\ell} + C^{\gamma_{E,I}^{i}, \kappa^{j}}_{\ell} + C^{\gamma_{E,I}^{i},\gamma_{E,I}^{j}}_{\ell}\, ,
\label{eq:cell_components}
\end{align}
\noindent noting that the terms involving lensing are all computed with the relevant matter auto- or cross-power spectrum, e.g.\  $P_{{\kappa^{j},\kappa^{j}}} = P_{\delta_m\delta_m}$ and $P_{\delta_{g}^{i},\delta_{g, \mu}^{j}}(k) = P_{\delta_{g}^{i},\kappa^{j}}(k) = P_{\delta_{g}^{i},\delta_{m}}(k)$.

The projection kernels for the four components of interest in this work are: 
\begin{align}
    W^{\delta_{g}}(\chi) &= p^{\delta_{g}}(z(\chi))E(\chi) \nonumber \\
    W^{\gamma_{E}}(\chi) &= p^{\gamma}(z(\chi))E(\chi) \\
    W^{\kappa}(\chi) &= \frac{3}{2}\Omega_{m,0}H_{0}^2(1+z(\chi))\int_{\chi}^{\infty} d\chi^{\prime} g^{i}(z(\chi), z(\chi^{\prime})) \nonumber \\    
    W^{\delta_{g,\mu}}(\chi) &= 2(\alpha_{\mu} - 1)W^{\kappa}(\chi) \nonumber
\end{align}
\noindent where $p^{\delta_{g}}(z)$ and $p^{\gamma}(z)$ are the lens and source galaxy redshift distributions, respectively. Furthermore, the lensing efficiency is given by 
\begin{equation}
    g^{i}(z, z^{\prime}) = \frac{\chi(z)(\chi(z^{\prime}) - \chi(z))}{\chi(z^{\prime})} \ p(z(\chi^{\prime}))E(\chi^{\prime})\, ,
\end{equation}
\noindent where $p(z(\chi^{\prime}))$ is the source redshift distribution for $W^{\kappa}(\chi)$ and the lens redshift distribution for $W^{\delta_{g,\mu}}(\chi)$. Additionally, the lens magnification kernel, $W^{\delta_{g,\mu}}(\chi)$, is modulated by the response of the lens density to magnification at the selection boundary of the sample\footnote{We note that magnification coefficients are often quoted in terms of the logarithmic derivative of $n$ with respect to limiting magnitude, $s_{\mu}=(1/n)\,dn/dm|_{m_{\rm lim}}$.  This implies $\alpha_{\mu}=(5/2)s_{\mu}$ so that $2(\alpha_\mu-1)=5s_\mu-2$.}:
\begin{align}
\label{eq:magnification_coeff}
    \frac{1}{n^{i}}\frac{d n^{i}}{d\kappa} = 2 (\alpha_{\mu} - 1)\quad .
\end{align}
Values of $\alpha_\mu$ for each lens sample are given in Table \ref{tab:lens_info}.

\subsection{Lagrangian Perturbation Theory and Hybrid Effective Field Theory}

\subsubsection{Galaxy and Matter Density}

We now specify the three dimensional power spectra used to compute these angular contributions. In particular, for the galaxy density and galaxy-matter power spectra we make use of hybrid effective field theory (HEFT) \cite{Modi2019,Kokron2021,Hadzhiyska2021,Zennaro2022,Pellejero2023,DeRose2023,Nicola2024,Shiferaw2025,Zhou2025,Bartlett2026}, assuming a second-order Lagrangian bias expansion\footnote{Truncating the expansion at second order is adequate for the scales and types of galaxies of interest in this work.  For tests and extension beyond quadratic order see the HEFT references cited above.} for the galaxy overdensity \cite{Desjacques2018}:
\begin{equation}
 \label{eq:bias_exp}
    \delta_g\left[\delta(\bq)\right] \approx  1 + b_1 \delta_0 + b_2 \left(\delta^2_0 - \langle \delta^2_0\rangle\right) +  b_s \left(s^2_0 - \langle s^2_0 \rangle\right)  +b_{\nabla^2}\nabla^2 \delta_0 + \epsilon \nonumber \, ,
\end{equation}
\noindent where $\bq$ is the Lagrangian coordinate, whose dependence we have suppressed on the RHS, $\delta_0$ and $s_0$ are the linear density and tidal field respectively, and $\epsilon$ is an uncorrelated stochastic contribution which we assume to have a white spectrum ($P_\epsilon=$const). The $b_{\mo_i}$ coefficients in front of each operator are the so-called galaxy bias coefficients, describing the unknown proportionality between the operators and the Lagrangian galaxy density field. We then use $N$-body simulations to compute fully non-linear displacements $\Psi(\bq, \tau)$, which we use to advect the Lagrangian fields to their late-time coordinates.
The power spectrum of the galaxy overdensity field can then be expressed as a sum over the advected operators, $\mo_i(\bx)$:
\begin{align}
    \label{eq:pgg}
    P_{\delta_{g}\delta_{g}}(k) &= \sum_{\mo_i,\mo_j\in \delta_g} b_{\mo_i}b_{\mo_j}P_{\mo_i\mo_j}(k) + P_\epsilon \\ 
    \label{eq:pgm}
    P_{\delta_g\delta_{m}}(k) &= \sum_{\mo_i\in \delta_g} b_{\mo_i}P_{1\mo_i}(k) \\
    \label{eq:pmm}    
    P_{\delta_m\delta_m}(k) &= \left( 1 - a_2  \frac{(\tilde{R}\, k/k_{\rm bar})^2}{1+(\tilde{R}\, k/k_{\rm bar})^2} \right) P_{11}(k)\, ,    
\end{align}
\noindent We make use of the \texttt{Aemulus $\nu$} emulator \cite{DeRose2023} for these basis spectra, $P_{\mo_i\mo_j}(k)$, and we remind the reader that $P_{11}(k)$ is just the standard total matter power spectrum in the absence of hydrodynamics. We include all operators to second order, neglecting cubic bias as it is degenerate with the $k^2$ counterterm at one-loop order.

Importantly, we also include counterterms in our prediction for the galaxy and matter power spectra and these can account for deviations from the baryon-free prediction provided by our simulation-based emulator. For $P_{\delta_m\delta_m}(k)$ we have re-summed terms beyond the $\nabla^2\delta$ counterterm into a Pad\'{e} approximant, with a dimensionless free parameter $\tilde{R}$. This model has been shown to fit a wide range of hydrodynamical simulations to $k=1\kMpc$ \cite{Chen2024b,DeRose2025,Bartlett2026}. Throughout this work we set $k_{\rm bar}=1.25\kMpc$, which implies a $\sim 5\%$ contribution to $P_{\delta_m\delta_m}(k)$ at $k=0.4\kMpc$ for $a_2=0.5$ and $\tilde{R}=1$.

We also note that our model takes as input re-parameterized bias parameters:
\begin{equation}
    \tilde{b}_{O^{(n)}} = b_{O^{(n)}} \left( \sigma_8(z)/\sigma_{8,\rm fid} \right)^n,\, .
    \label{eqn:tilde_bias}
\end{equation}
\noindent This reparameterization is not particularly important for this work, as it is primarily used to remove posterior projection effects \cite{Pandey2020,Chen2024b,Maus24b,Tsedrik2025} when fitting these models to data using MCMC techniques, rather than the Fisher forecasting we will be concerned with.

\subsubsection{Intrinsic Alignments}

For the trace-free component of the galaxy shape field, $g_{ij}$, we can write down a similar second-order Lagrangian expansion to that in Equation~\ref{eq:bias_exp} \cite{Vlah2021,Chen2024a}:
\begin{align}
    g_{ij}[ & L_{ij}(\bq)] \approx  \, c_{1} s_{ij} + c_{\delta1}\delta s_{ij} + c_{t} t_{ij}  + c_{2} \mathrm{TF}\{s^2\}_{ij} + \alpha_{s}\nabla^2 s_{ij} + \epsilon_{ij},
  \label{eq:shape_exp}
\end{align}
where the $c_{\mo_i}$ coefficients are galaxy shape bias parameters, analogous to the galaxy density bias parameters $b_{\mo_i}$. This expansion is written in terms of the Lagrangian shear tensor $L_{ij}(\bq)=\partial_{i}\Psi_{j}$, and $\delta=-\mathrm{tr}(L_{ij}^{(1)})$, $s_{ij}$ and $t_{ij}$ are the linear Lagrangian density, and first and second order tidal shear tensors. We will adopt the helicity basis \cite{Vlah2020} with angular momentum $\ell=0$ for the galaxy densities and $\ell=2$ for the galaxy shapes.
In this basis, only power spectra where both fields have the same helicity, $m$, are non-zero by spherical symmetry about $\hat{k}$. The Lagrangian shape operators can be advected\footnote{In fact the process of advection generates non-zero higher order terms even if they vanish initially, in a way similar to the ``coevolution'' relations for galaxy bias \cite{Schmitz2018}.} similarly to the bias operators, leading to expansions of the three-dimensional power spectra 
\begin{align}
\label{eq:shape_shape_basis}
    P_{\ell\ell^{\prime}}^{(m)}(k) = \sum_{\mo_{i}, \mo_{j} \in g } c_{\mo_i}c_{\mo_j}P_{\ell\ell^{\prime},\mo_{i}\mo_{j}}^{(m)}(k)\\
\label{eq:density_shape_basis}
    P_{02}^{(0)}(k) = \sum_{\mo_{i}\in \delta, \mo_{j} \in g } b_{\mo_i}c_{\mo_j}P_{02,\mo_{i}\mo_{j}}^{(0)}(k)
\end{align}
where Eq.~\ref{eq:density_shape_basis} holds for any scalar $\delta$, but in this case we will be concerned with galaxy densities $\delta_g$, and matter overdensities, $\delta_m$. In the case of the matter density-cross spectrum, all density biases, $b_{\mo_i}$ are set to zero in Eq.~(\ref{eq:density_shape_basis}). We can then write the $E$-mode auto-power spectra and the galaxy or matter density-E-mode cross-power spectrum that enter Eq.~(\ref{eq:limber}) as \cite{Vlah2021}
\begin{equation}
    P_{\gamma^i_{E,I} \gamma^j_{E,I}}(k) = \frac38 P_{22,ij}^{(0)}(k) + \frac{1}{8} P_{22,ij}^{(2)}(k)
    \quad , \quad
    P_{\delta^{i} \gamma^i_{E,I}}(k) = \sqrt{\frac{3}{8}} P_{02, ij}^{(0)}(k) \ .
\end{equation}
To compute the spectra entering into Eqs.~(\ref{eq:shape_shape_basis}, \ref{eq:density_shape_basis}) we use the Lagrangian perturbation theory code, \texttt{spinosaurus} \cite{Chen2024a}, specifically the \texttt{ShapeShapeCorrelators} method for $P_{22,\mo_{i}\mo_{j}}^{(m)}$ and the \texttt{DensityShapeCorrelators} method for $P_{02,\mo_{i}\mo_{j}}^{(0)}$.  Note that due to a different convention employed, the outputs from \texttt{DensityShapeCorrelators} need to be multiplied by $\sqrt{3/2}$ to match the conventions adopted here (i.e.\ those from ref.~\cite{Vlah2020}) for $P_{02,ij}$.  See Appendix \ref{app:IA_conventions} for further discussion and comparison of IA conventions.

Although we assume a perturbative model for the IA contributions, our scale cuts will be set by the significantly larger contributions from the weak lensing signal. Given this, we will probe scales that exceed $k_{NL}$ in the perturbative IA model where we expect these IA predictions to break down. Since the cosmological information is saturated at large scales, we anticipate that this IA treatment will be sufficient, though future, tighter constraints on baryonic feedback may change that conclusion

Of equal or greater importance to assumptions about scale dependence of the IA contributions, defined by the operators that we have included above, is the form of the redshift dependence of the intrinsic alignment parameters. Many previous analyses have assumed that the IA parameters do not vary with redshift \cite{Asgari2020} or vary as a simple power-law in redshift \cite{Abbot2021,Krause2021}. Recent investigations have assumed independent IA amplitudes per source bin, allowing for uncertainties on these amplitudes proportional to the fraction of red galaxies in each bin Refs.~\cite{McCullough2024,Wright2025} or inversely proportional to redshift overlap with spectroscopic samples \cite{Bigwood2026}. The potential for these assumptions to bias cosmological inference has been pointed out by a number of authors \cite{Chen2024a,Blot2025}, and we will revisit this question below.

In particular, if one is agnostic to the form of IA redshift evolution in a cosmic shear analysis, then the IA signal is highly degenerate with $S_8$, with only weak degeneracy breaking coming from the fact that the II and GI terms have different redshift dependence, and from priors that enforce the IA contribution to be small compared to the noise. The IA contributions to galaxy-galaxy lensing signals are directly proportional to the overlap between source and lens redshift distributions, making the IA-$S_8$ degeneracy essentially perfect for source-lens bin pairs with significant redshift overlap. While this is a nuisance for \twotimestwo\ analyses, effectively removing the cosmological constraining power of these IA dominated bins, for \threetimestwo\ analyses this is a boon as it allows one to measure IA signals quite well for IA dominated source-lens bin pairs, and if the lens redshift distributions have large enough overlap with the source redshift distributions then this can cleanly break the IA-$S_8$ degeneracy in the cosmic shear data, so long as sufficient freedom is allowed in the IA redshift evolution model to account for the level of redshift evolution in these signals in the data. 

As we will show below for both Steel and Gold samples, \twotimestwo\ analyses are constraining enough to self-calibrate even relatively complex models for IA redshift evolution, while strong assumptions about IA redshift evolution only serve to potentially bias constraints. In order to thoroughly investigate this behavior we allow each IA parameter, $c_{\mo}$, to evolve as a spline in redshift, independently for each source bin:
\begin{equation}
    c_{\mo}^i(z) = \sum_{j=0}^{N} c_{\mo}^{ij}\ W\left( \frac{z - z_{\rm min}}{\Delta} - j \right),
    \label{eqn:spline}
\end{equation}
\noindent where $i$ indexes over source bins, $\Delta$ is the pre-determined node spacing defining the smoothness of the redshift dependence and the spline covers points between $z_{\rm min}$ and $z_{\rm max} = z_{\rm min} + N \Delta$. For simplicity we assume that the splines are linear between nodes, i.e.\ $W(x) = \text{max}(0,1-|x|)$. For our fiducial forecasts, we assume $z_{\rm min}=0$ and $z_{\rm max}=3.0$ with $\Delta=0.5$. We will investigate the dependence of our forecasts on $\Delta$ below, using the data-driven scenario for IA redshift evolution described in Appendix~\ref{app:ia_z_dependence}, comparing to the assumption of a constant amplitude (scaled by linear growth) per source bin as assumed in, e.g., \cite{Wright2025,Bigwood2026}. 

Again we note that our model takes in a reparameterized form of the IA parameters:
\begin{equation}
    \tilde{c}_{\mo^{(n)}} = c_{\mo^{(n)}} \left( \sigma_8(z)/\sigma_{8,\rm fid} \right)^n,\, ,
    \label{eq:tilde_ia}
\end{equation}
\noindent similar to what was done for the bias parameters above. More importantly, we do not include the conventional prefactor of $\rho_c C_1\Omega_m$, which for $\Omega_m=0.31$ is equal to $0.0043$. Thus, a value of $A_1=1$ using typical normalization conventions is equal to $c_s=0.0086$, where the additional factor of two comes from the difference between normalizing our basis spectra with respect to the the three-dimensional rather than projected alignment amplitude. See additional discussion in App.~\ref{app:IA_conventions}.

\subsection{Lensing Counter Terms}

Although we use a simulation based model for the matter power spectrum in our predictions for lensing auto-spectra (e.g.\ \ckk), our model for baryonic effects on the matter power spectrum (Equation~\ref{eq:pmm}) is only accurate at the $1\%$ level to $k\simeq 1\kMpc$ \cite{DeRose2025}, and beyond $k\sim1\kMpc$ it is likely that baryonic effects reach the $>10\%$ level and become even more uncertain. Instead of making stringent scale cuts to remove angular scales contaminated by $k>1\kMpc$, which inevitably will remove information from $k<1\kMpc$, we can instead marginalize over these highly uncertain scales in a model-agnostic manner by simply expanding the $k>1\kMpc$ contributions to \cgg\ as \cite{Chen2024b,DeRose2025}
\begin{align}
    C_{\ell,\rm UV}^{\kappa^i\kappa^j} &= \int_{\Lambda}^{\infty} \frac{dk}{(\ell + 1/2)}  W^{\kappa^i}\Big(\frac{\ell + 1/2}{k}\Big) W^{\kappa,j}\Big(\frac{\ell + 1/2}{k}\Big) P_{\rm \delta_m\delta_m}^{\rm UV}\left(k,\chi = \frac{\ell + 1/2}{k}\right) \nonumber \\
    &= \left(\frac32 \Omega_m H_0^2\right)^2 \sum_{N=2}^\infty (\ell + 1/2)^{N-1} \sum_{n+m+o=N} w_n^i w_m^j \sigma^{\rm UV}_{N,o} \label{eq:cl_uv}
\end{align}
\noindent where $\Lambda=1\kMpc$ is the scale beyond which we assume we cannot model accurately. The first line in Equation~\ref{eq:cl_uv} above simply changes the Limber integration variable to $k =(\ell + 1/2)/\chi$, and the second line expands $W^{\kappa}$ about $\chi=0$
\begin{equation}
    W^{\kappa^i}(\chi) = \frac32 \Omega_m H_0^2 \sum_{n=1}^\infty w_n^{i} \chi^n
\end{equation}
in addition to defining the lensing counter terms $\sigma_{N,o}$ by expanding $P_{\rm \delta_m\delta_m}^{\rm UV}(k)$ about $\chi=0$ and integrating over $k$ as
\begin{equation}
\label{eqn:sigmas}
    \sigma^{\rm UV}_{N,o} = \frac{1}{o!} \int dk\ k^{-N}\ \left. \frac{d^o P^{\rm UV}_{\delta_m\delta_m}}{d\chi^o}\right|_{\chi=0}\, .
\end{equation} 
The $w_n$ coefficients can be computed analytically for a given cosmology and source redshift distribution, leaving $\sigma_{N,o}$ as free parameters that encode our uncertainties on $P^{\rm UV}_{\delta_m\delta_m}(k)$. We then take
\begin{equation}
    \sigma^{\rm UV}_{N,o} = \sigma^{\rm UV, \rm fid}_{N,o}f^{\rm UV}_{N,o}
    \label{eqn:lct_param}
\end{equation}
where $\sigma^{\rm UV, \rm fid}_{N,o}$ are computed by plugging in our fiducial \texttt{Aemulus $\nu$} matter power spectrum model into Equation~\ref{eqn:sigmas}, and $f^{\rm UV}_{N,o}$ then characterizes the fractional deviation of $\sigma^{\rm UV}_{N,o}$ away from its CDM prediction. In this work, we truncate the expansion in Eq.~\ref{eq:cl_uv} at $N=4$, which was shown to be the optimal choice in \cite{DeRose2025} given the convergence radius of this expansion, thus adding six free parameters to our model.

\subsection{Observational Systematics}
\label{sec:obs_sys}
In this work, we make relatively simple assumptions about the form that shear systematics take, allowing for a free multiplicative bias parameter for each source bin, motivated by the significant progress in this field over the last decade \cite{Huff2017,Sheldon2017,Sheldon2020,Li2025}:
\begin{align}
\label{eqn:cgg_calib}
    \cgg &\to (1 + m_{i})(1 + m_j) \cgg\\
\label{eqn:cdg_calib}
    \cdg &\to (1 + m_j) \cdg \,.
\end{align}

On the other hand, there are still major uncertainties related to how redshift distributions for the faint galaxy samples used in weak lensing studies will be calibrated. In light of these uncertainties, we adopt a flexible parameterization of our source galaxy redshift distributions, allowing for four degrees of freedom per redshift bin: a shift and stretch
of the core redshift distribution (Eq.~\ref{eq:nz_core}):
\begin{equation}
    p^{\gamma\prime}_{i,\rm core}(z|\Delta z_{\rm core}^i,\sigma_{z,\rm core}^i) = p^{\gamma}_{i,\rm core}(z + \Delta z_{\rm core}^i | \sigma_{z,\rm core}^i)\,,
\label{eqn:delta_z_source}    
\end{equation}
and a free outlier fraction and mean redshift of the outlier distribution:
\begin{equation}
    p^{\gamma\prime}_{i,\rm out}(z|\Delta z_{\rm out}^i) = p^{\gamma}_{i,\rm out}(z + \Delta z_{\rm out}^i)\,.
\label{eqn:delta_z_source_out}    
\end{equation}
This is very similar to the parameterization of Ref.~\cite{Zhang2026}, with the additional freedom permitted by $\Delta z_{\rm out}^i$ to shift the location of the outliers (though for simplicity we still keep shape of the outlier distribution fixed).  The total source galaxy redshift distribution model is thus given by
\begin{equation}
\label{eq:nz_uncertainty}
    p^{\gamma^{\prime}}_i(z) = (1 - f_{{\rm out}}^i)\,p^{\gamma\prime}_{i,\rm core}(z|\Delta z_{\rm core}^i,\sigma_{z,\rm core}^i) + f_{{\rm out}}^i\,p^{\gamma\prime}_{i,\rm out}(z|\Delta z_{\rm out}^i)\,.
\end{equation}
Table~\ref{tab:source_nz_priors} outlines our assumptions about the means and uncertainties on these parameters that we assume in our analysis. Note that for Steel prior widths, unlike for Gold, we have not included the usual $1+z$ factor, as our goal is to directly calibrate the redshift distributions of all bins to the level given by our fiducial priors.

\begin{table*}[t!]
    \centering
    \begin{tabular}{|c|c|c|c|}
         \hline
         \hline
          & Steel & Gold (Pessimistic) & Gold (Optimistic) \\
         \hline
         $\bar{n}$ [arcmin$^{-2}$] & 5.0 & 27.7 & 27.7\\
         $\sigma_e$ & 0.26 & 0.26 & 0.26 \\         
         $n_{\rm bins}$ & 4 & 5 & 5 \\         
         $\langle \sigma_{z,\rm core}^{i}\rangle$ & $0.05(1+z)$ & $\{6.5, 8.3, 9.8, 12, 19\}\times 10^{-2}$ & $\{6.5, 8.3, 9.8, 12, 19\}\times 10^{-2}$ \\
         $\langle f_{\rm out}^{i} \rangle$ & 0.01 & 0.15 & 0.15\\
         $\sigma(n^{i}(z))$  & $0.005$ & $0.01(1+z)$ & $0.001(1+z)$ \\
         \hline 
    \end{tabular}
    \caption{Summary of sample characteristics and priors on Steel and Gold source galaxy redshift distribution uncertainties. We list number densities, $\bar{n}$, shape noise, $\sigma_e$, and number of redshift bins, $n_{\rm bins}$, fiducial core redshift distribution widths, $\langle \sigma_{z,\rm core}^{i}\rangle$, and outlier fractions, $\langle f_{\rm out}^i\rangle$. The priors on the redshift parameters, $\sigma(\Delta z_{\rm core}^i)$, $\sigma(\Delta z_{\rm out}^i)$, $\sigma(\sigma_{z,\rm core}^{i})$, and $\sigma(f_{\rm out}^{i})$ are the same for a given sample, and listed in the row indicated as $\sigma(n^{i}(z))$.}.
    \label{tab:source_nz_priors}
\end{table*}

\section{Priors and Forecasting Framework}
\label{sec:framework}

\begin{table*}
\caption{Parameters and priors. Here $\mathcal{U}(a,b)$ indicates a uniform distribution on $[a,b]$ while $\mathcal{N}(\mu,\sigma^2)$ indicates a normal distribution with mean $\mu$ and variance $\sigma^2$.}
\begin{center}
\begin{tabular}{| c  c  c  |}
\hline
\hline
Parameter & Prior & Reference\\  
\hline 
\multicolumn{3}{|c|}{{\bf Cosmology}} \\
$\omega_c$  &  $\mathcal{U}$($0.08, 0.16$) &  \\ 
$\omega_b$ & $\mathcal{U}$($0.0173, 0.0272$) &  \\
$A_s$ &  $\mathcal{U}$($1.1\times 10^{-9},3.1\times 10^{-9}$) & \\
$n_s$ & $\mathcal{U}$($0.93, 1.01$) & Sec. \ref{sec:framework} \\
$h$  & $\mathcal{U}$($0.52, 0.82$) & \\
$\sum \, m_{\nu}$  & 0.06 eV & \\

\hline
\multicolumn{3}{|c|}{{\bf Lens galaxy bias}} \\
$(\sigma_8(z)/\sigma_{8,\rm fid})(1 + b_{1}^{i}) $  & $\mathcal{U}$ ($0.5, 3.5^2$) & Eq. \ref{eq:bias_exp}-\ref{eq:pgm}\\
$(\sigma_8(z)/\sigma_{8,\rm fid})^2\,b_{2}^{i} $  & $\mathcal{N}$ ($0, 1^2$) & Eq. \ref{eq:bias_exp}-\ref{eq:pgm}\\
$(\sigma_8(z)/\sigma_{8,\rm fid})^2\,b_{s}^{i} $  & $\mathcal{N}$ ($0, 1^2$) & Eq. \ref{eq:bias_exp}-\ref{eq:pgm} \\
$2k_{\rm max}^2 b_{\nabla^2 a}^{i}/(1+b^{i}_1) $ & $\mathcal{N}$ ($0, 0.2^2$) & Eq. \ref{eq:bias_exp},\ref{eq:pgg}\\
$2k_{\rm max}^2 b_{\nabla^2 \times}^{i}/(1+b^{i}_1) $ & $\mathcal{N}$ ($0, 0.2^2$) & Eq. \ref{eq:bias_exp},\ref{eq:pgm}\\
$\textrm{SN}^{i}$ & $\mathcal{N}(\textrm{Table \ref{tab:lens_info}}, 30\%)$ & Eq. \ref{eq:bias_exp},\ref{eq:pgg}\\

\multicolumn{3}{|c|}{{\bf Lensing counterterms}} \\
$f^{UV}_{N,o}$   & $\mathcal{N}$ ($0,0.4^2$) & Eq. \ref{eqn:sigmas}\, , \ref{eqn:lct_param}\\
\hline
\multicolumn{3}{|c|}{{\bf Baryonic effects}} \\
$a_2^i$   & $\mathcal{N}$ ($0,0.5^2$) & Eq. \ref{eq:pmm}\, , \ref{eqn:spline}\\
$\tilde{R}^i$   & $\mathcal{N}$ ($0,1^2$) & Eq. \ref{eq:pmm}\, , \ref{eqn:spline}\\
\hline
\hline
\multicolumn{3}{|c|}{{\bf Intrinsic alignment}} \\
$(\sigma_8(z)/\sigma_{8,\rm fid})c_{1}^{ij}$   & $\mathcal{N}$ ($0,0.04^2$) & Eq. \ref{eq:shape_exp}\\
$(\sigma_8(z)/\sigma_{8,\rm fid})^2\,c_{2}^{ij}$   & $\mathcal{N}$ ($0,0.04^2$) & Eq. \ref{eq:shape_exp}\\
$(\sigma_8(z)/\sigma_{8,\rm fid})^2\,c_{\delta 1}^{ij}$   & $\mathcal{N}$ ($0,0.04^2$) & Eq. \ref{eq:shape_exp}\\
$(\sigma_8(z)/\sigma_{8,\rm fid})^2\,c_{t}^{ij}$   & $\mathcal{N}$ ($0,0.04^2$) & Eq. \ref{eq:shape_exp}\\
$(\sigma_8(z)/\sigma_{8,\rm fid})^2\,\alpha_{s}^{ij}$   & $\mathcal{N}$ ($0,1.8^2$) & Eq. \ref{eq:shape_exp}\\

\hline 
\multicolumn{3}{|c|}{{\bf Magnification \& shear calibration}} \\
$\alpha^{i}_{\rm \mu}$ & $\mathcal{U}(\textrm{Table \ref{tab:lens_info}}\pm 0.1)$ & Eq. \ref{eq:magnification_coeff}\\
$m^{i}$ & $\mathcal{N} (0,0.003^2)$ & Eq. \ref{eqn:cgg_calib} \\
\hline
\multicolumn{3}{|c|}{{\bf Source \photoz\ }} \\
$\Delta z^{i}_{\rm core}$  & $\mathcal{N}$ ($0$, \text{Table \ref{tab:source_nz_priors}}) & Eq. \ref{eq:nz_uncertainty}  \\
$\Delta z^{i}_{\rm out}$  & $\mathcal{N}$ ($0$, \text{Table \ref{tab:source_nz_priors}}) & Eq. \ref{eq:nz_uncertainty}  \\
$\sigma_{z,\rm core}^{i}$  & $\mathcal{N}$ (\text{Table \ref{tab:source_nz_priors}}) & Eq. \ref{eq:nz_uncertainty}  \\
$f_{\rm out}^{i}$  & $\mathcal{N}$ (\text{Table \ref{tab:source_nz_priors}}) & Eq. \ref{eq:nz_uncertainty}  \\
\hline
\end{tabular}
\end{center}
\label{tab:params}
\end{table*}

For our cosmological forecasts and Fisher bias calculations, we produce Fisher matrices working at the two-point level:
\begin{equation}
    F_{\alpha\beta} = \frac12 \partial_\alpha\mu_kC^{-1}_{kl} \partial_\beta \mu_l + C_{\alpha\beta, \rm prior}^{-1} 
\end{equation}
where $\mu_k$ is the model prediction for a given bandpower, derivatives are taken with respect to the model parameters via automatic differentiation using the \texttt{gholax}\footnote{\href{https://github.com/j-dr/gholax}{https://github.com/j-dr/gholax}} code. For galaxy clustering and galaxy-galaxy lensing we assume the $\ell_{\rm max}$ given for each lens bin in Table~\ref{tab:lens_info}. For cosmic shear, we follow the procedure outlined in Ref.~\cite{DeRose2025} for determining cosmic shear scale cuts, removing all bandpowers where the LCT expansion using a cutoff scale of $\Lambda=1\kMpc$ is inaccurate at the $1\%$ level. The resulting $\ell_{\rm max}$ values for each cosmic shear bin pair are listed in Table~\ref{tab:ell_max_shear}. Ref.~\cite{DeRose2025} demonstrated that these provide significantly more constraining power than naive $k=1\kMpc$ scale cuts as defined in e.g.\  Ref.~\cite{Doux2022} by mitigating the impact of mode mixing between scales with $k<\Lambda$ and $k>\Lambda$. For simplicity, we compute these scale cuts for the Gold sample redshift distributions and use them for both Gold and Steel analyses. This very moderately underestimates the $\ell_{\rm max}$ that can be used for Steel cosmic shear analyses, due to the compactness of the Steel redshift distributions compared with Gold, but we have checked that this does not significantly impact our results. 

\begin{table}
    \centering
    \caption{Maximum multipole $\ell_{\rm max}$ used for each cosmic shear bin pair $(i,j)$, used for both Steel and Gold samples.} 
    \label{tab:ell_max_shear}
    \begin{tabular}{c|ccccc}
        \hline\hline
        $i \backslash j$ & 0 & 1 & 2 & 3 & 4 \\
        \hline
        0 & 116 & 169 & 173 & 180 & 182 \\
        1 &     & 789 & 866 & 925 & 977 \\
        2 &     &     & 977 & 1058 & 1134 \\
        3 &     &     &     & 1166 & 1271 \\
        4 &     &     &     &      & 1415 \\
        \hline
    \end{tabular}
\end{table}

We assume $C_{\rm prior}^{-1}$, the inverse covariance matrix of the parameter priors, is diagonal and make the disconnected approximation for $C^{-1}$, the inverse data covariance matrix, such that 
\begin{equation}
[C_{kl}]^{XY} = \frac{1}{f_{\rm sky}\Delta\ell_k(2\ell_k + 1)\delta_{kl}^{K}}
\begin{cases} 
      2C^{\delta_g^{i}\delta_g^{j}}_{k}C^{\delta_g^{i}\delta_g^{j}}_{l} & X=\delta_g^i\delta_g^i\,, Y =\delta_g^j\delta_g^j \\
      2C^{\delta_g^{i}\delta_g^{j}}_{k} C^{\delta_g^{i}\gamma_{E}^{l}}_{l} & X=\delta_g^i\delta_g^i\,, Y=\delta_g^j\gamma_{E}^l \\
      (C^{\delta_g^{i}\delta_g^{m}}_{k} C^{\gamma_{E}^j\gamma_{E}^{n}}_{l} + C^{\delta_g^{i}\gamma_E^{n}}_{k} C^{\delta_g^{m}\gamma_E^{j}}_{l})  & X=\delta_g^i\gamma_{E}^j\,, Y=\delta_g^m\gamma_{E}^n \\ 
      (C^{\delta_g^{i}\gamma_E^{m}}_{k} C^{\gamma_E^{j}\gamma_{E}^{n}}_{l} + C^{\delta_g^{i}\gamma_E^{n}}_{k} C^{\gamma_E^{j}\gamma_{E}^{m}}_{l})  & X=\delta_g^i\gamma_{E}^j\,, Y=\gamma_E^m\gamma_{E}^n \\       
      (C^{\gamma_E^{i}\gamma_E^{m}}_{k} C^{\gamma_E^{j}\gamma_{E}^{n}}_{l} + C^{\gamma_E^{i}\gamma_E^{n}}_{k} C^{\gamma_E^{j}\gamma_{E}^{m}}_{l})  & X=\gamma_E^i\gamma_{E}^j\,, Y=\gamma_E^m\gamma_{E}^n \, ,
\end{cases}
\end{equation}
where for auto-power spectra we include noise terms in the spectra $N^{\delta_g^i\delta_g^i} = 1/\bar{n}_i$ and $N^{\gamma_E^i\gamma_E^i} = \sigma_e^2/\bar{n}_i$. The angular number densities entering into these terms are those measured for DESI galaxies for $\delta_g$ shown in Table~\ref{tab:lens_info} converted to sr$^{-1}$, and $1/4$ of the total source galaxy density for the Steel sample, and $1/5$ for the Gold sample, corresponding to using four or five source bins with equal surface densities respectively. Although the disconnected covariance approximation may break down in detail, any breakdown of this assumption will lead to more correlation between small scale modes, thus reducing the cosmological constraining power of these scales and decreasing the returns of reducing noise terms.

We also include Fisher bias calculations for a few scenarios, where parameter biases are calculated as 
\begin{equation}
\Delta\theta_\alpha = (F^{-1})_{\alpha\beta} (\partial \mu_j/ \partial \theta_\beta)(C^{-1})_{jk}\, \Delta \mu_k \, ,
\end{equation}
where $\Delta \mu_k$ is the systematic contamination for which we wish to calculate parameter biases.

The priors for our analysis are listed in Table~\ref{tab:params}. In order to calculate $\sigma_8$, we produce 10000 samples from the Gaussian Fisher posteriors and use a neural network emulator to predict $\sigma_8$ given the sampled cosmological parameters, which is trained in the \texttt{Aemulus $\nu$} parameter space \cite{DeRose2023}. While the training space is wide compared to our tightest constraints, it does truncate the tails of our least constraining forecasts for relatively unconstrained parameters such as $\omega_b$, placing an implicit flat prior on the cosmological parameters as listed in Table~\ref{tab:params}.

\section{Forecast Constraints}
\label{sec:constraints}

We now present forecasts for the $S_8$ constraining power of DESI-LSST combined analyses as a function of source galaxy number density, intrinsic alignment, redshift distribution, and baryonic uncertainties.  We break these results up into a series of subsections detailing how different assumptions about the samples or modeling affect the forecasts.

\subsection{Source galaxy number density}

To begin with, we investigate how the $S_8$ constraining power of a combined DESI-LSST analysis depends on the source galaxy density of the Steel sample. We compare this to two different LSST Y10 Gold-like sample scenarios, one with optimistic redshift calibration assumptions and one with pessimistic assumptions. As discussed in Section~\ref{sec:obs_sys}, the pessimistic Gold sample assumptions take $\sigma(\Delta z_{\rm core}^i) =0.01(1+z)$ and $\sigma(\sigma_{z,\rm core}^i) =0.01(1+z)$, comparable to (but notably better than) the levels of calibration achieved in Stage III surveys. The optimistic scenario assumes the level of redshift calibration assumed in the DESC SRD, i.e.\ $\sigma(\Delta z_{\rm core}^i) =0.001(1+z)$ and $\sigma(\sigma_{z,\rm core}^i) =0.001(1+z)$ even though a path that can achieve such a calibration for the full density of the Gold sample is currently unknown, particularly for the widths of the core distributions \cite{dAssignies2025}. We assume that we can calibrate the redshift distributions of the Steel sample to $\sigma(\langle z\rangle)=0.005$, which we will motivate in Section~\ref{sec:feasibility}. Additional details regarding redshift calibration assumptions can be found in Section~\ref{sec:lsst} and Table~\ref{tab:source_nz_priors}.

\begin{figure}
    \centering
    \includegraphics[width=\linewidth]{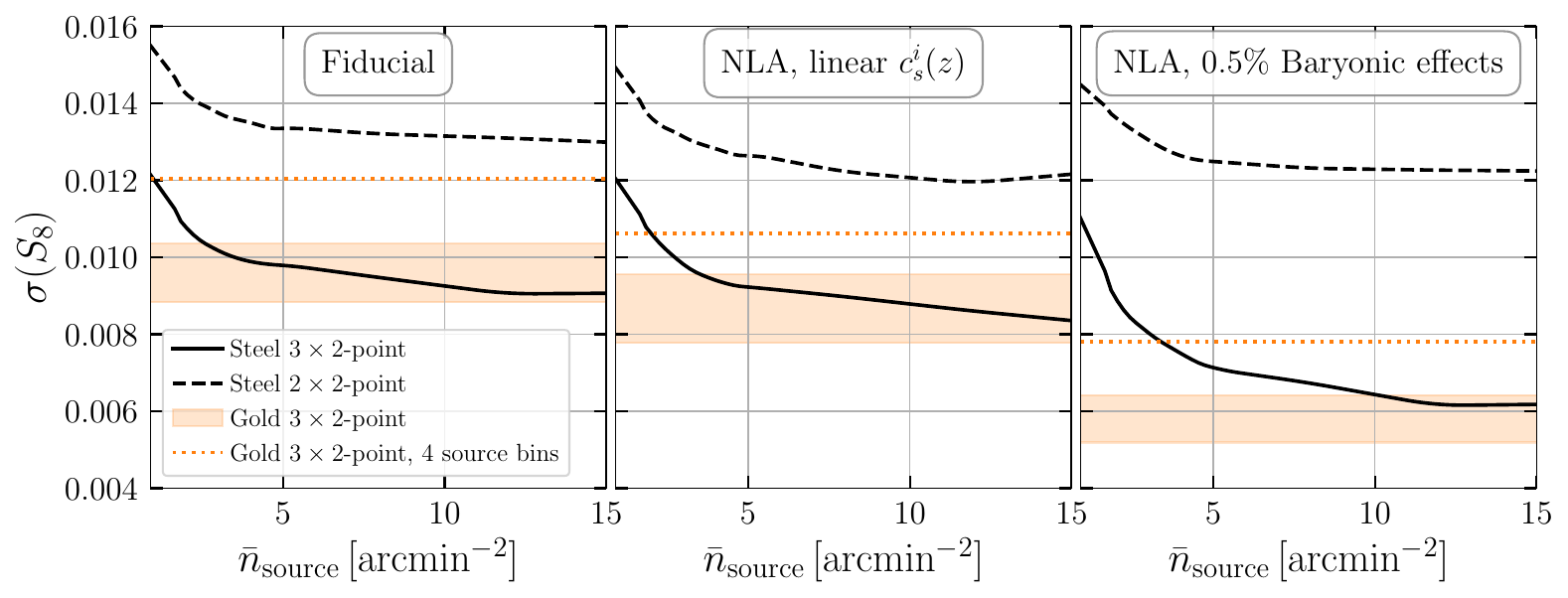}
    \caption{Impact of source number density on $S_8$ constraining power under various modeling assumptions for the Steel redshift distributions (black) for \threetimestwo\ (solid), and \twotimestwo\ (dashed) analyses compared to \threetimestwo\ constraints from the Gold sample (orange band; with $\bar{n}_{\rm source}=27.7\,\mathrm{arcmin}^{-2}$).  The upper end of the Gold sample constraining power corresponds to stage III levels of redshift calibration while the lower limit corresponds to the levels required by the SRD, analogous to $\sigma(\Delta z)=0.001$. The dashed orange line shows the most pessimistic Gold scenario where we must discard the highest redshift source bin. The left panel shows forecasts for our fiducial modeling assumptions. The middle panel simplifies the IA treatment to use the NLA model with linear redshift evolution per source bin. The right panel shows a case with this simplified IA model and the impact of baryons reduced by a factor of 10, such that there are only $0.5\%$ baryonic effects at $k=0.4\kMpc$.}
    \label{fig:Steel_3x2_forecast}
\end{figure}

Figure~\ref{fig:Steel_3x2_forecast} shows the results of these forecasts for three different assumptions about IA and baryon modeling. The left panel assumes our fiducial model for IA redshift evolution, truncating the IA expansion at second order, and setting baryon parameters $\sigma(a_{2}^{i})=0.5$ and $\sigma(\tilde{R}^{i})=1$, equivalent to a $\sim 5\%$ error on the matter power spectrum at $k=0.4\kMpc$. The black solid and dashed lines show \threetimestwo\ and \twotimestwo\ analyses using the Steel sample as sources, while varying the total density of the Steel sample according to the $x$-axis. The orange bands span the optimistic and pessimistic redshift calibration scenarios for the Gold sample, fixing the density to the fiducial Gold sample density of 27.7 arcmin$^{-2}$. Although the parametric form assumed for our redshift distributions is more flexible than what most other forecasts have investigated, there is still a risk that it underestimates uncertainties, particularly in the form of outlier distributions, for the hightest redshift sources. If this is the case then the highest redshift source galaxies may have to be discarded entirely, and we indicate such a Gold sample scenario as the dotted orange lines. For our fiducial modeling choices, we see that the constraining power of the \twotimestwo\ analysis saturates at densities of $\simeq 2\, \rm arcmin^{-2}$, while \threetimestwo\ analyses approach significantly diminishing returns at $\simeq 5\, \rm arcmin^{-2}$. Both of these are dramatically less than the density of the Gold sample. 

In the middle panel, we show the same results but now simplifying the IA treatment to use the commonly used non-linear linear alignment (NLA) model \cite{Hirata2004,Bridle2008} with linear redshift evolution in $c_s^{i}$ for each source bin. The observed behavior is largely the same as for our fiducial modeling choices shown on the left panel, with the main differences being that the constraining power of the pessimistic Gold scenario improves by $\sim 5\%$, and that the Steel sample constraining power saturates at somewhat larger number densities. The rightmost panel then shows the same IA modeling assumptions, but shrinks the priors on the baryon-effect parameters by a factor of ten, such that baryons now impact the matter power spectrum at the $0.5\%$ level at $k=0.4\kMpc$. In this case the constraining power at all number densities for all samples significantly improves, and improvements with larger number densities become more significant. Thus we see that baryonic effects limit the utility of higher source number densities more than IAs for a \threetimestwo\ point analysis. In the first two cases, the Steel sample with a number density of $\simeq 5\, \rm arcmin^{-2}$ is more constraining than the Gold sample with pessimistic redshift calibration assumptions. In the highly optimistic scenario that baryonic effects are constrained to $0.5\%$ level at $k=0.4\kMpc$, the Steel sample at this density and the Gold sample are competitive.  We explore our sensitivity to baryonic effects and scale cuts in more detail below. For the rest of this work, we will assume a source density of $5\, \rm arcmin^{-2}$ for the Steel sample, which is towards the optimistic end of the range of values we motivate in more detail in Section~\ref{sec:feasibility}. 

\subsection{Intrinsic alignment self-calibration}
\label{sec:ia_forecasts}

Now we investigate the interesting result that our forecast constraining power for Steel and Gold samples are relatively insensitive to IA modeling assumptions. This is perhaps surprising for the Gold sample, given the significant overlap between source and lens redshift distributions for most of the source-lens bin pairs, due to our assumptions about the tails of the Gold redshift distributions. As a simple test, we compare the lensing--only signal-to-noise ratio (SNR) in source-lens bin pairs for which IAs are a negligible contribution to the signal-to-noise in the full galaxy-galaxy lensing data vector, where negligible is quantified as having an IA contribution that is less than 10\% of the statistical error assuming $c_s=0.008$ (analogous to $A_1\sim1$). For both Gold and Steel samples we find that the difference in the SNR of the uncontaminated GGL signal to the full GGL signal is $<5\%$, i.e.\ the signal in bins uncontaminated by IA far exceeds the signal in the contaminated bins. This is because the bins that have small IA contributions are also the bins whose lensing signal is largest. For the Gold sample, the increased lensing signal contributed by higher redshift sources outweighs the increase in overlap between source and lens redshift distributions, mitigating the latter as a source of additional systematic error.

\begin{figure}
    \centering
    \includegraphics[width=\linewidth]{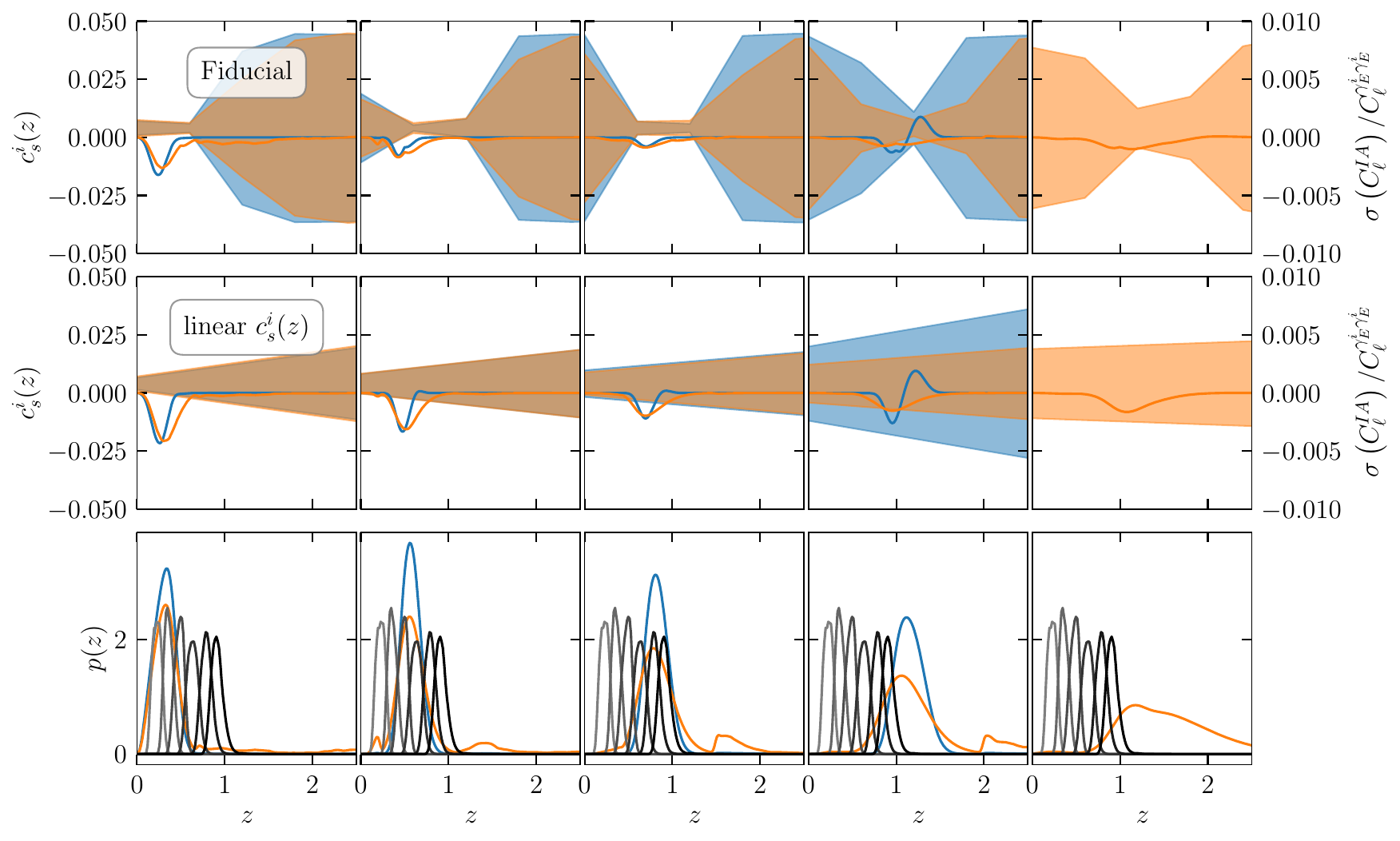}
    \caption{Illustration of the IA self-constraining power of \threetimestwo\ analyses even in the presence of flexible IA redshift evolution. The shaded regions in the top two rows show constraints on $c_s^{i}(z)$ for Steel (blue) and Gold (orange) for each source bin in the various columns. The lines show the uncertainty in the fractional IA contribution to the total signal in the cosmic shear auto-power spectrum at $\ell=200$ as a function of redshift. The top row shows the case of our fiducial IA redshift evolution model with a correlation length of $\Delta z=0.5$.  The middle row assumes a linear spline per source bin, i.e. $\Delta z=3.0$. The bottom row shows the source galaxy redshift distributions (blue and orange for Steel and Gold respectively) as well as the lens redshift distributions (grayscale). The \twotimestwo\ data tightly constrains the IA amplitudes in both IA redshift evolution scenarios which mitigates IA contamination to cosmic shear.}
    \label{fig:ia_self_calib}
\end{figure}

The above argument explains why the \twotimestwo\ constraints are not significantly impacted by our IA modeling choices, but does not explain why the \threetimestwo\ constraints are similarly unaffected. In order to investigate the latter phenomenon, we now examine the ability of the \threetimestwo\ analysis to `self calibrate' IA contamination. This is illustrated in Figure~\ref{fig:ia_self_calib}, which shows the posterior on the $c_s^{i}(z)$ for each source bin in the various columns for a \threetimestwo\ analysis, for two levels of IA redshift evolution complexity. In the top we show our fiducial model for $c_s^{i}(z)$, setting $\Delta z_{c_s}=0.5$, while in the middle row we show results for linear evolution of $c_s^{i}(z)$, i.e.\ $\Delta z_{c_s}=3.0$. The bottom row shows the source redshift distributions for Gold and Steel in orange and blue, with lens redshift distributions over-plotted in black. The shaded regions in the top two rows show the posteriors on $c_s^{i}(z)$ for each bin, with a $y$-axis scale shown on the left hand side of the plot, while the lines show the fractional uncertainty in the IA contribution to the cosmic shear auto power spectrum of that bin at $\ell=200$ as a function of redshift. The $y$-axis on the right hand side shows the scale of these contributions. Integrating these lines over redshift gives the uncertainty in the fractional IA contribution to these spectra, i.e. the fractional systematic uncertainty in cosmic shear due to IAs. This ranges from more than $10\%$ for the first redshift bins in the Steel and Gold samples, to $\sim 0.5\%$ for the highest redshift Steel sample bin and $2.5\%$ for the highest redshift Gold bin. 

Generically, we see that the $c_s^{i}(z)$ constraints are tightest in the redshift range where the IA contributions peak. This is driven by the fact that the lens redshift distributions have significant overlap with the source bins in the redshift range where the IA contributions to cosmic shear become largest. Thus the galaxy-galaxy lensing signal for bin pairs with significant source-lens overlap serve to constrain the IA contributions to the cosmic shear data, thus significantly boosting the power of cosmic shear when included in the \threetimestwo\ data combination.

Given these results, one natural question to ask is how the choice of IA spline spacing, $\Delta z_{c_s}$, which can be roughly interpreted as the correlation length assumed in our IA redshift evolution model, impacts the constraining power in our forecasts. Furthermore, we wish to investigate the level of flexibility that is warranted in our IA redshift evolution model given a plausible scenario for IA redshift evolution. A model with $\Delta z_{c_s}=0.5$ is significantly more conservative than what has been assumed in any \threetimestwo\ analysis to date, but it is not maximally agnostic in the sense argued for in Ref.~\cite{Chen2024b}, in that this correlation length is significantly larger than the width of a given lens bin and so multiple lens-source bin pairs contribute to constraining the IA amplitude at a given spline node. The potential for assumptions about IA redshift evolution to bias cosmological inference in a DES Y3-like scenario was explored in Ref.~\cite{Chen2024b} using a simplified model. Here we wish to investigate this question further using a data-driven model for LSST Gold and Steel-sample scenarios. 

To construct this scenario we measure the luminosity and red-fraction redshift evolution of Gold- and Steel-like samples from the HSC data, and then use measured IA amplitude-luminosity relations for red and blue galaxies from \cite{Joachimi2012,Siegel2026a,Girones2026} to construct scenarios for $c_s^{i}(z)$ for the Gold and Steel samples independently. Additional details regarding the assumptions in this model are described in Appendix~\ref{app:IA_variations}. We refer to this scenario as ``Biased IA$(z)$''.  This model was then used to produce contaminated \threetimestwo\ data vectors, $\mu_c$, which we fit with our \threetimestwo\ model assuming the NLA model with either constant or spline redshift evolution, producing best fit $\hat{\mu}_{\Delta z_{c_s}}$ models. The residuals of these best fit models $\Delta \mu_{\Delta z_{c_s}}=\mu_c-\hat{\mu}_{\Delta z_{c_s}}$ were used as input to our Fisher bias calculations.

\begin{figure}
    \centering
    \includegraphics[width=\linewidth]{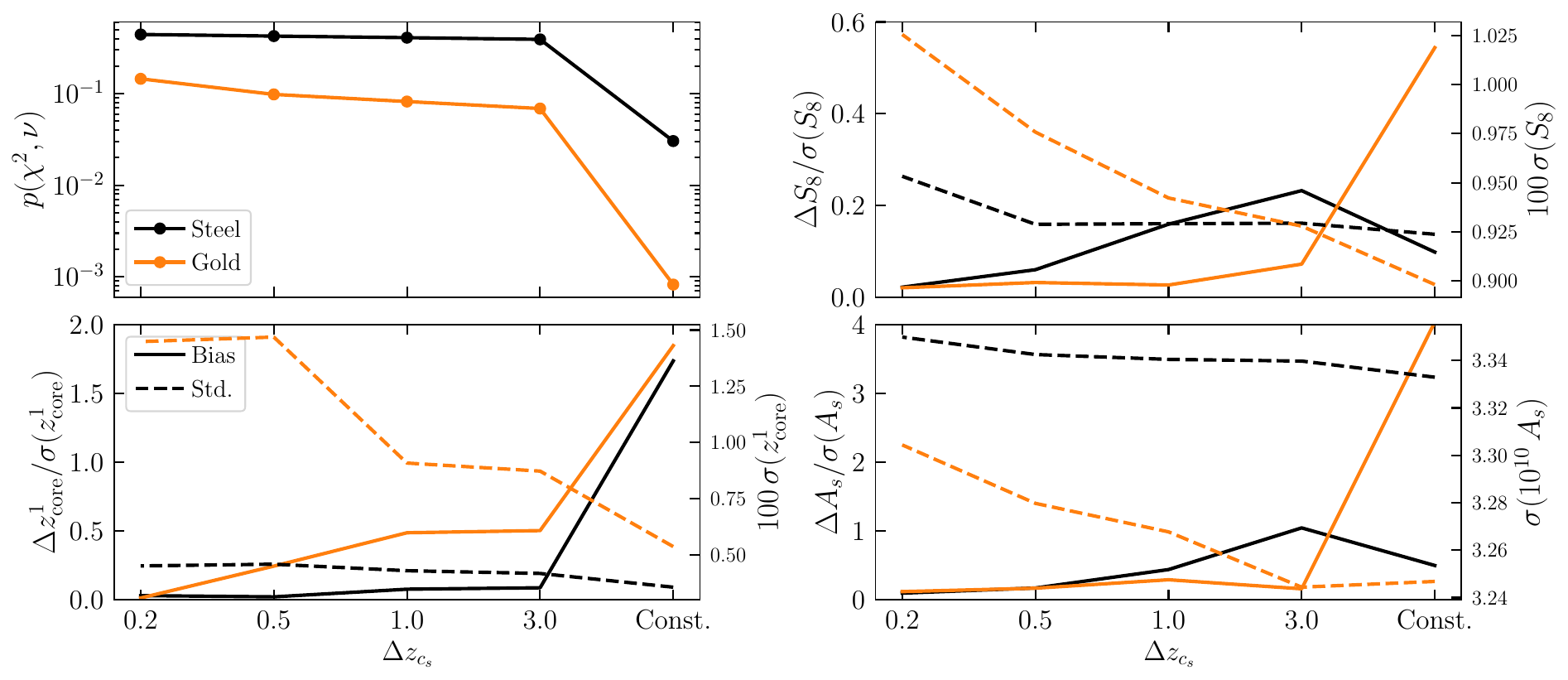} 
    \caption{$p$-values for $\nu$ degrees of freedom (top left) and Fisher bias calculations in units of the uncertainty for $S_8$ (top right), $\Delta z_{\rm core}^1$ (lower left) and $A_s$ (lower right) as a function of $\Delta z_{c_s}$, the spline node spacing for $c_s$. $p$-values are for fitting the ``Biased IA$(z)$'' scenario for Steel (black) and pessimistic Gold (orange) samples, varying only the IA sector. Fisher uncertainty forecasts are shown as dashed lines with the right $y$-axis of the three `bias' panels, with Fisher bias values the solid lines with the left $y$-axis. Generically, we see that models that only allow for a constant amplitude per source bin, denoted as ``Const.'' on the $x$-axis, provide very poor fits to the ``Biased IA$(z)$'' model, and that biases decrease quickly as model complexity increases.}
    \label{fig:ia_bias}
\end{figure}

Figure~\ref{fig:ia_bias} shows the results of these calculations, where the black lines show results for a Steel sample with our fiducial $5\, \rm arcmin^{-2}$ density, and orange lines indicate results for the Gold sample with pessimistic redshift calibration assumptions throughout. The top-left panel shows the $p$-value of the $\chi^2$ for the fits of each model to the ``Biased IA$(z)$'' scenario; $\nu$ is the number of degrees of freedom in the fit. Here we have assumed that the $\chi^2=\tilde{\chi}^2+\nu$ where $\tilde{\chi}^2$ is the $\chi^2$ value obtained by fitting our \threetimestwo\ model with spline IA evolution to the ``Biased IA$(z)$'' and $\nu$ is the expected contribution from statistical noise, i.e. $\langle\chi\rangle^2=\nu$. We see that models that only allow for a constant amplitude per source bin, denoted as ``Const.'' on the $x$-axis, provide very poor fits to the ``Biased IA$(z)$'' model and would yield $p$-values assuming mean noise fluctuations of $p=0.03$ and $p=8\times10^{-4}$ for Steel and Gold respectively. Decreasing to $\Delta z_{c_s}=3$, adding one extra parameter per source bin, increases these $p$-values to $p=0.4$ for the Steel sample, i.e.\ very close to the expectation for the case of no bias, $p=0.5$. The same model also increases the $p$-values for Gold, with values remaining around $p=0.1$ for all $\Delta z_{c_s}$ values considered.

The top-right panel is similar to the top-left panel, but shows Fisher bias and uncertainty calculations for $S_8$ as solid and dashed lines respectively, with the scale indicated on the left and right $y$-axes showing biases and uncertainties respectively. The Fisher bias values, $\Delta S_8$, are quoted as fractions of the uncertainty, $\sigma(S_8)$. In the Gold scenario, we see a relatively large bias of $\Delta S_8=0.5\,\sigma$ for the one-constant-per-source-bin IA model, which rapidly decreases as we decrease $\Delta z_{c_s}$. The Steel sample does not show significant biases for any value of $\Delta z_{c_s}$. We see that the constraining power of the Steel sample analysis is effectively insensitive to $\Delta z_{c_s}$. This is due in part to the cleaner separation of IA and lensing signals in the \twotimestwo\ data vector due to the assumed compactness of the Steel sample, as well as to the greater statistical error in Steel measurements. The constraining power of the Gold sample analysis gradually degrades as we decrease $\Delta z_{c_s}$, reaching a $9\%$ degradation with respect to the constant amplitude per bin scenario for $\Delta z_{c_s}=0.2$. 

This conclusion is substantially different from that found in Ref.~\cite{Chen2024b}, where significant constraining power was lost by adding IA model complexity. The reason for this difference in conclusions is largely due to the relatively low redshift, and broad nature of the DES-Y3 source bins. This again emphasizes the significant gains that can be derived from compact $n(z)$. Similar results in the context of the NLA-$z$ model compared to a constant NLA model for all source bins were shown in Ref.~\cite{Blot2025}. 

The bottom-left and bottom-right panels have the same format as the top-right panel, but highlight the behavior of the most biased redshift calibration, $\Delta z_{\rm core}^1$, and cosmological parameter, $A_s$. The $\Delta z_{\rm core}^1$ panel shows similar behavior to the $S_8$ panel, except now with $>1\sigma$ biases for the constant $c_s$ model for both Steel and Gold. We also see that the ability to self-calibrate the redshift parameters degrades as we increase the IA model complexity. Although we have not plotted it here, similar behavior is seen in the IA sector, with $>1\sigma$ biases for the constant NLA per source bin case, and degradation of self-calibration ability as model complexity increases. For the constant NLA per bin case, we see that $A_s$ is biased at approximately $4\sigma$ and $0.5\sigma$ for Gold and Steel respectively. The bias in Gold decreases quickly to under $0.5\sigma$ for $\Delta z_{c_s}\leq 3$, while the bias in the Steel sample analysis persists at the $\sim1-2\sigma$ level until $\Delta z_{c_s}=0.5$. Given the very moderate decreases in constraining power that are observed as $\Delta z_{c_s}$ decreases, we advocate for a relatively conservative choice of $\Delta z_{c_s}=0.5$.

\begin{figure}
    \centering
    \includegraphics[width=\linewidth]{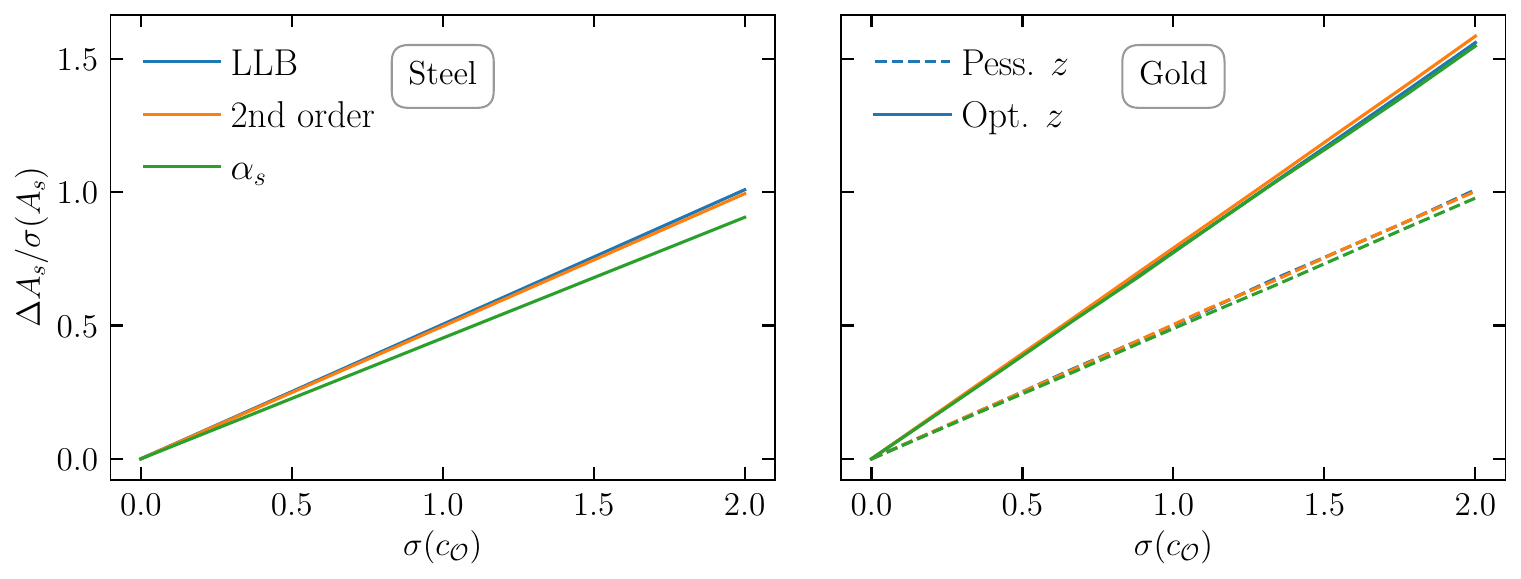} 
    \caption{Bias (in units of the uncertainty) in $A_s$ from mis-modeling IA as a function of the size of the non-linear IA contribution (as a fraction of our fiducial priors $\sigma(c_{\mathcal{O}})$). The data are fit with NLA with a constant amplitude per source bin, and the non-linear IA amplitudes are also assumed to be constant in redshift. We show results for three different contamination models: second order Lagrangian IAs (orange), a local Lagrangian bias (blue), and a model with just a counterterm in addition to $c_s$ (green). Left and right panels show Steel and Gold samples, the latter with pessimistic (dashed) and optimistic (solid) redshift calibration.}
    \label{fig:nlia_bias}
\end{figure}

In addition to the redshift evolution of intrinsic alignments, another potential source of bias is the scale dependence of IAs beyond the NLA model. Our fiducial IA model includes the full second-order Lagrangian expansion (Eq.~\ref{eq:shape_exp}), parameterized by $c_2$, $c_{\delta 1}$, $c_t$, and $\alpha_s$ in addition to the linear alignment amplitude $c_s$. However, the NLA model, which retains only $c_s$, is commonly used in practice. To investigate the potential for non-linear IA contributions to bias cosmological inference when fitting with NLA, we perform Fisher bias calculations where we contaminate our \threetimestwo\ data vectors with non-linear IA contributions of varying amplitude and fit with an NLA model assuming a constant $c_s$ per source bin. For this exercise, we assume that the non-linear IA parameters are constant with redshift. Figure~\ref{fig:nlia_bias} shows the resulting bias on $A_s$, quoted as a fraction of the forecast uncertainty, as a function of the true non-linear IA amplitude expressed as a fraction of our fiducial priors $\sigma(c_{\mathcal{O}})$. We consider three contamination scenarios: the full second-order Lagrangian expansion, the local Lagrangian bias (LLB) assumption, i.e. freeing $c_{\delta 1}$ and $\alpha_s$ in addition to $c_s$, and a model including only the counterterm $\alpha_s$ in addition to $c_s$. The left and right panels show results for the Steel and Gold samples respectively. For both samples, biases grow as the non-linear IA contributions increase, and exceed $\sim 0.5\sigma$ biases around $\sigma(c_{\mathcal{O}})=1$. The Gold sample with pessimistic redshift calibration and Steel samples are comparably susceptible to these biases while Gold with optimistic redshift calibration is significantly more susceptible. The biases from parameters other than $\alpha_s$ are relatively negligible. Importantly, these results motivate the use of at least the LLB model, as the minimal unbiased, self-consistent model. This model has also been shown to deliver comparable goodness of fit to the full one-loop Lagrangian IA model on simulated halo samples \cite{Chen2024a}, and similar models accurately predict three-point alignments in simulations \cite{Vedder2026}, further advocating for its sufficiency for near-term datasets.

\begin{figure}
    \centering
    \includegraphics[width=\linewidth]{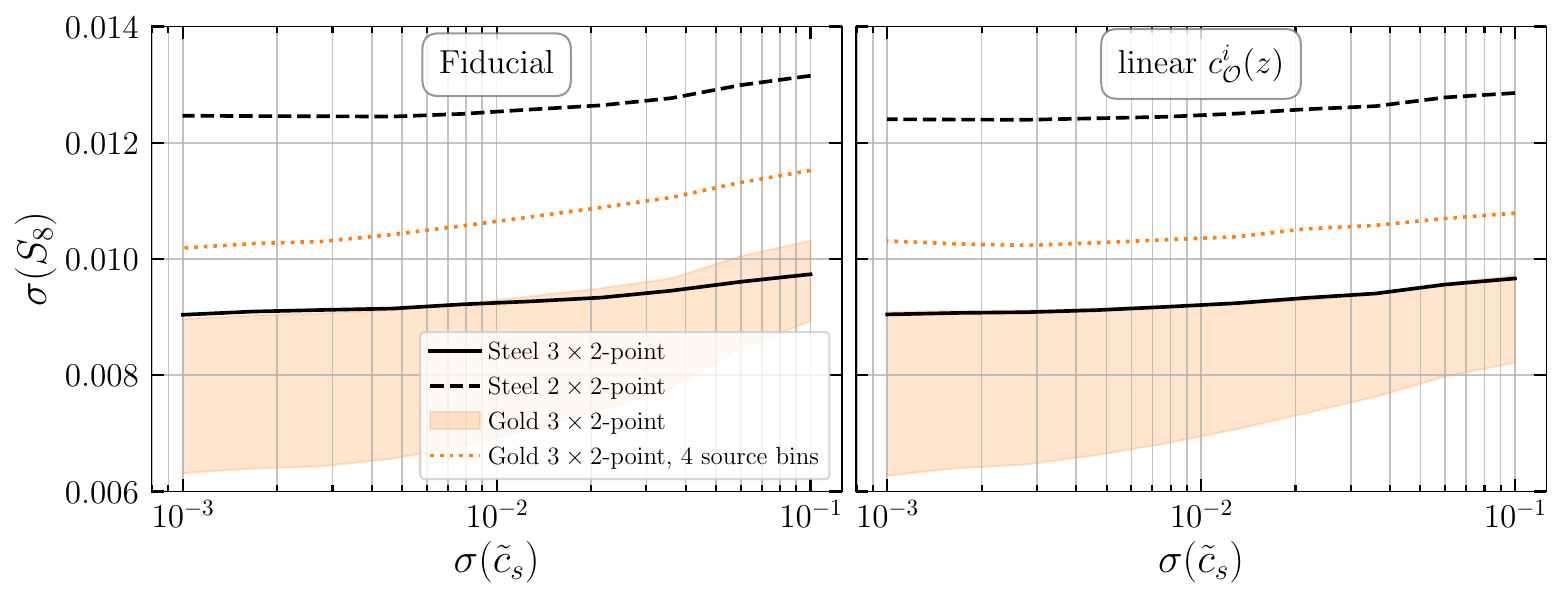}    
    \caption{({\it Left}) Error on $S_8$ as a function of the IA prior width under our fiducial modeling assumptions for the Steel sample \threetimestwo\ (solid black) and \twotimestwo\ (dashed black) analyses compared to a Gold sample \threetimestwo\ analysis (orange). Dotted orange lines show a Gold-like analysis with the highest redshift source bin discarded. Steel forecasts assume our fiducial number density of $5\, \rm arcmin^{-2}$. As before, the range of Gold constraints spans stage III photo-z priors ($\sigma(\Delta z_s)=0.01$) to SRD-like priors ($\sigma(\Delta z_s)=0.001$) with all other \photoz\ priors scaled by the same factor. The $\tilde{c}_s$ in the $x$-axis label is a proxy; we scale all IA priors by the same factor that the $\tilde{c}_s$ priors are scaled by. {(\it Right}) Same as left, but now assuming that all IA parameters evolve linearly with redshift.}
    \label{fig:s8_v_cs}
\end{figure}

As a final investigation of the sensitivity of our analyses to IAs, we show how $\sigma(S_8)$ varies as a function of our priors on IA amplitudes in Figure~\ref{fig:s8_v_cs}. The left and right hand sides of this figure show results for our fiducial modeling choices with $\Delta z_{c_s}=0.5$ and for linear spline $c_{\mathcal{O}}^{i}(z)$ with $\Delta z_{c_s}=3.0$, respectively. The $x$-axis shows the prior width on $\tilde{c}_s$, but we shrink the priors on all IA parameters by the same fraction with respect to their fiducial settings listed in Table~\ref{tab:params}. We see that for \twotimestwo\ and \threetimestwo\ analyses, shown in dashed and solid lines respectively, the Steel sample results are effectively insensitive to the IA priors. The Gold sample with pessimistic redshift calibration assumptions, shown as the top of the filled orange contour, is similarly insensitive to IA prior widths. Once redshifts have been calibrated to the levels used in our optimistic Gold forecasts, then improvements are seen in $\sigma(S_8)$ as the IA priors are reduced as shown by the lower edge of the orange contour. Thus, by externally calibrating the IA amplitudes of the Gold sample to $\sigma(c_s)=0.002$, up to $30\%$ ($15\%$) improvements in $\sigma(S_8)$ are possible for our fiducial modeling choices (linear $c^{i}_{\mo}(z))$, although constraining $c_s$ to this level of accuracy, roughly equivalent to $\sigma(A_1)=0.125$, would require an extremely expensive spectroscopic program. We forecast the density and area of spectroscopy that would be required to reach this level of calibration in Appendix~\ref{app:direct_ias}.

\subsection{Redshift uncertainties}
\label{sec:z_forecasts}

\begin{figure}
    \centering
    \includegraphics[width=\linewidth]{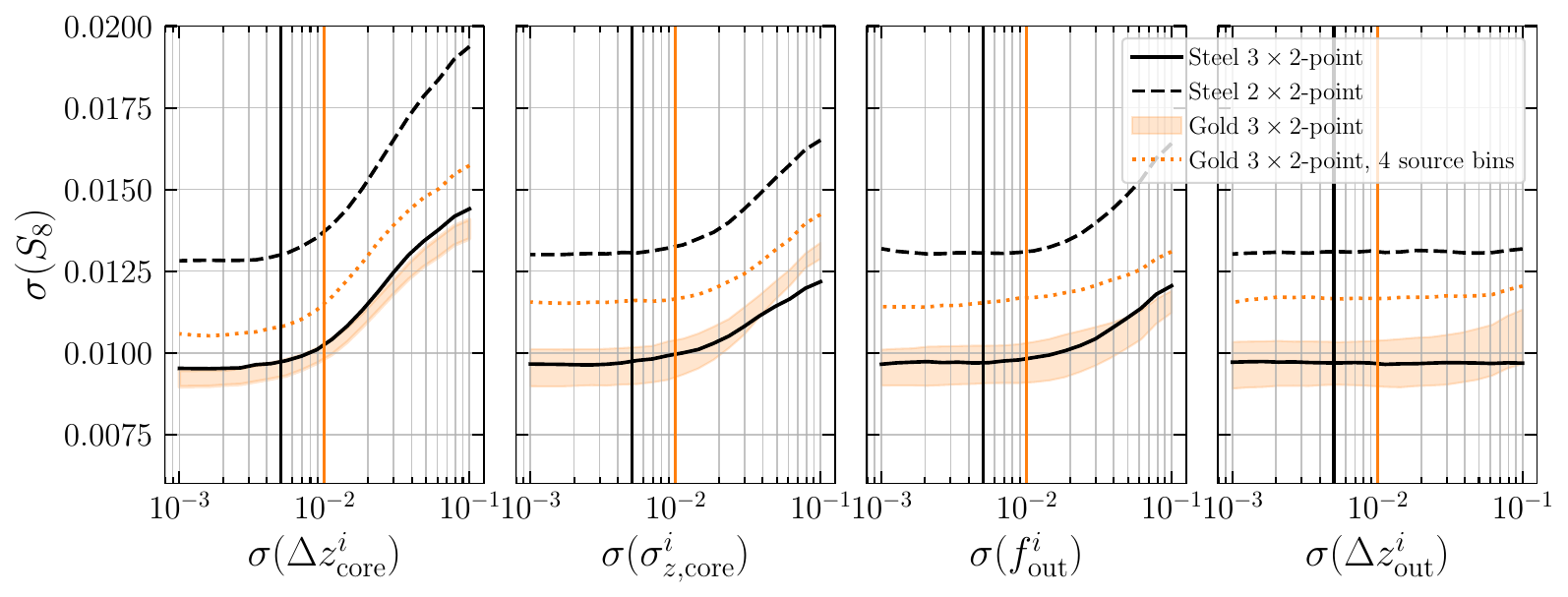}
    \caption{Error on $S_8$ as a function of the assumed uncertainty on the mean (left) and width (middle left) of the core redshift distributions as well as the outlier fraction (middle right) and uncertainty on the mean redshift of the outlier distribution (right). All forecasts assume our fiducial modeling assumptions for the Steel sample \threetimestwo\ (solid black) and \twotimestwo\ (dashed black) analyses compared to a Gold sample \threetimestwo\ analysis (orange). We also note that for Gold, the we include a factor of $(1+z)$ in our redshift uncertainties in the two left panels, multiplying the values on the $x$-axis, while for Steel we do not include this factor given our goal of directly spectroscopically calibrating this sample. Steel forecasts assume our fiducial number density of $5\, \rm arcmin^{-2}$. The upper (lower) boundary of the Gold constraints fixes all other redshift nuisance parameter uncertainties to the pessimistic (optimistic) redshift optimization scenario as described in the text. The dotted gold line removes the highest redshift Gold source bin on top of the pessimistic Gold assumptions.}
    \label{fig:pz_forecast}
\end{figure}

Now we turn to an investigation of how the constraining power of the Steel and Gold sample analyses depend on redshift uncertainties. Figure~\ref{fig:pz_forecast} is analogous to Figure~\ref{fig:s8_v_cs}, now showing how $\sigma(S_8)$ depends on the priors on parameters describing source galaxy redshift uncertainties. The panels show $\sigma(S_8)$ as a function of uncertainties in the mean redshift of the core distributions, $\Delta z_{\rm core}^{i}$, width of the core distributions, $\sigma_{z,\rm core}^{i}$, outlier fraction, $f_{\rm out}^i$ and the mean redshift of the outlier distributions, $\Delta z_{\rm out}^i$ from left to right. Here the upper (lower) edge of the Gold contours fix all redshift uncertainty parameters to their pessimistic (optimistic) fiducial values shown in Table~\ref{tab:source_nz_priors} except for the parameter being varied on the $x$-axis. These panels are also roughly ordered by importance of the parameter on $\sigma(S_8)$, with uncertainties in $\Delta z_{\rm core}^{i}$ influencing the $S_8$ uncertainty most strongly, and uncertainties in $\Delta z_{\rm out}^i$ having the smallest impact. For the $\sigma(\Delta z_{\rm core}^{i})$ panel, the Gold priors are scaled by $1+z$, while the Steel priors are not, given our intent to spectroscopically calibrate all bins of the Steel sample.

For both the Steel and Gold samples, we see that the $S_8$ uncertainty is very sensitive to $\sigma(\Delta z_{\rm core}^{i})$ until $\sigma(\Delta z_{\rm core}^{i})=0.005$, which is what sets our requirement on how well these parameters should be determined with the Steel sample. The Gold sample is slightly more sensitive than the Steel sample to the widths of the core redshift distributions, although the constraining power of both Steel and Gold samples saturate below $\sigma_{z,\rm core}^{i}\sim0.01$. Both Gold and Steel sample analyses show very similar trends with $f_{\rm out}^i$, with a weak dependence above $\sigma(f_{\rm out}^{i})=0.01$. Neither sample shows a significant trend with $\sigma(\Delta z_{\rm out})$, although this may be due to our simplistic assumption about how uncertainties in the outlier distribution may manifest themselves.

\begin{figure}
    \centering
    \includegraphics[width=\linewidth]{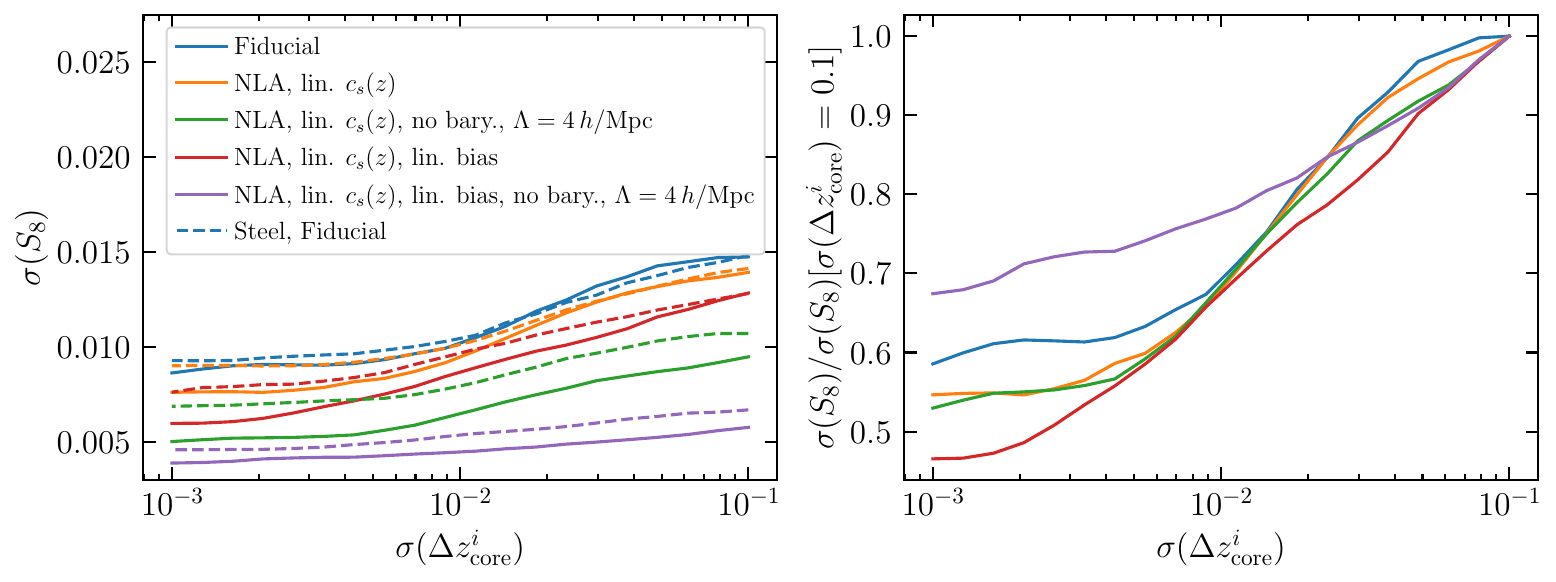}
    \caption{Impact of modeling analysis choices on the sensitivity of $S_8$ constraints to redshift calibration priors. ({\it Left}) Absolute constraints on $S_8$ as a function of $\sigma(\Delta z_{\rm core}^i)$. This is used as a proxy for all redshift parameters which are scaled by the same factors. ({\it Right}) Improvement in constraining power relative to that obtained with $\sigma(\Delta z_{\rm core}^i)=0.1$. Blue shows our fiducial analysis choices, orange uses NLA with linear $c_s(z)$ per source bin, green additionally fixes all baryonic parameters to zero and relaxes scale cuts to $\Lambda=4\kMpc$, and red and purple are the same as orange and green respectively but use linear galaxy bias to $k=0.3\kMpc$ instead of HEFT to $k=0.4\kMpc$. Solid (dashed) lines show Gold (Steel) sample constraints. Gold forecasts at $\sigma(\Delta z_{\rm core})=0.01$ (0.001) and Steel at $\sigma(\Delta z_{\rm core})=0.005$ correspond to the redshift uncertainty assumptions for these samples in the rest of this work. See text for discussion.}
    \label{fig:pz_assumptions}
\end{figure}

The saturation in constraining power at $\sigma(\Delta z_{\rm core}^{i})=0.005(1+z)$ and $\sigma_{z,\rm core}^{i}\sim0.01(1+z)$ that we find for the Gold sample is somewhat different than what has been reported in the past, for example in \cite{Mandelbaum2017}, where requirements on mean source redshifts were determined to be $\sigma(\Delta z^{i})=0.001(1+z)$ and $\sigma(\sigma_{z,\rm core}^{i})\sim0.003(1+z)$. In order to investigate the source of this difference, we examine our sensitivity to redshift calibration priors under a number of different modeling assumptions in Figure~\ref{fig:pz_assumptions}. The left panel shows absolute constraints on $S_8$ as a function of $\sigma(\Delta z_{\rm core}^i)$, as a proxy for all redshift parameters, which we shrink by the same factors as $\sigma(\Delta z_{\rm core}^i)$. The panel on the right shows the improvement in constraining power relative to the that obtained with $\sigma(\Delta z_{\rm core}^i)=0.1$. The blue line shows our fiducial analysis choices, while orange show an NLA analysis with linear $c_s(z)$ per source bin, green additionally fixes all baryonic parameters to zero and relaxes our scale cuts to $\Lambda=4\kMpc$. Red and purple are the same as orange and green but make the additional simplifying choice of using linear galaxy bias to $k=0.3\kMpc$ (at which scale  this model would be significantly biased as demonstrated in e.g.~\cite{Chen2024b}), similar to \cite{Mandelbaum2017}, instead of HEFT to $k=0.4\kMpc$. The solid lines show Gold sample constraints while dashed show the Steel sample.

First, we note the significant changes in constraining power as as we simplify our analysis choices. In particular, removing the impact of baryonic effects and fitting to $\Lambda=4\kMpc$ boosts the constraining power of our forecasts considerably. The purple line is the closest analysis in this work to what is considered in \cite{Mandelbaum2017}, and here we see that $\sigma(S_8)$ does indeed continue to decrease all the way down to $\sigma(\Delta z^{i})=0.001(1+z)$, although in this case the improvement over $\sigma(\Delta z_{\rm core}^i)=0.1$ is actually smallest. These forecasts suggest that it is actually our use of non-linear bias that softens our requirements on redshift calibration to the greatest extent. Blue and orange lines here correspond to the two different IA assumptions used throughout the rest of this work, and for these we see that Steel outperforms, or performs similarly to gold until $\sigma(\Delta z_{\rm core})<0.01(1+z)$, in agreement with the rest of results. We emphasize that the simplified constraints are shown here for comparison purposes to previous literature only. As argued throughout the rest of the text, our fiducial choices are what we believe is required to deliver unbiased constraints.

\subsection{Baryons and feedback}
\label{sec:baryon_forecasts}
\begin{figure}[h!]
    \centering
    \includegraphics[width=\linewidth]{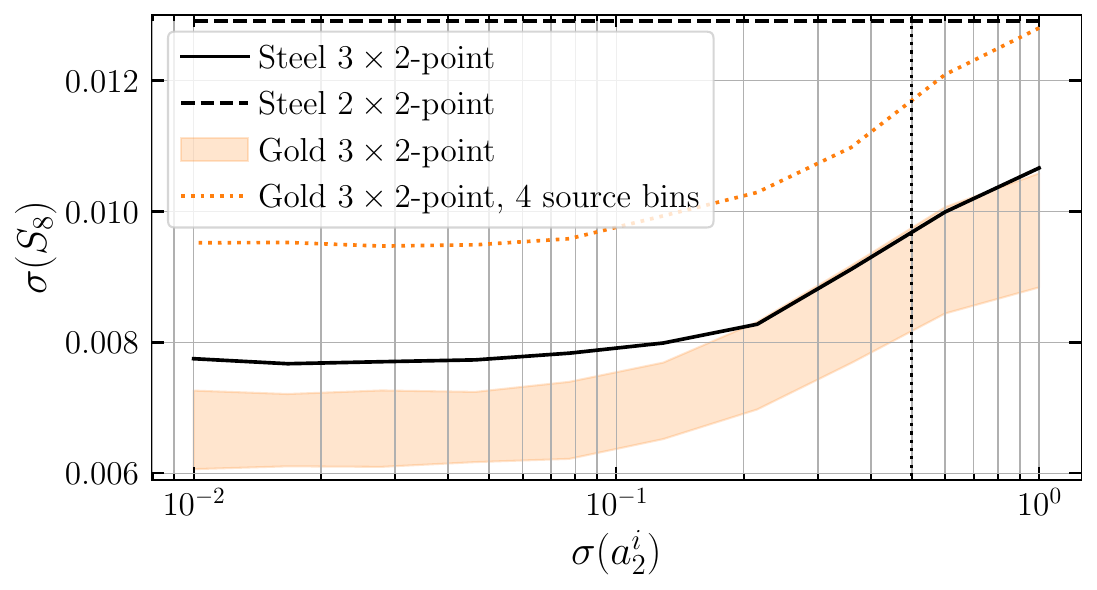}
    \caption{Error on $S_8$ as a function of the assumed uncertainty on parameters describing the effect of baryons on the matter power spectrum. The largest values in this plot correspond to $\sim 10\%$ baryonic feedback effects at $k=0.4\kMpc$ while the rightmost values correspond to $\sim 0.1\%$ contributions.}
    \label{fig:baryon_forecast}
\end{figure}

Finally, we examine the dependence of our forecasts to scale cuts and our priors on baryonic effects. Figure~\ref{fig:baryon_forecast} is analogous to Figure~\ref{fig:s8_v_cs}, except now we vary the priors on the baryonic effect parameters $a_2^{i}$ and $\tilde{R}^i$. The $x$-axis shows the prior width on $a_2^{i}$, and we vary the priors on $\tilde{R}^i$ by the same ratio with respect to the fiducial priors in Table~\ref{tab:params}, otherwise keeping our fiducial modeling choices fixed. We see that both Gold and Steel \threetimestwo\ analyses depend sensitively on our baryonic effect prior widths until $\sigma(a_2^i)\sim 0.1$ (and correspondingly $\sigma(\tilde{R}^i)\sim 0.2$), which corresponds to an uncertainty on the impact of baryons on the matter power spectrum of $\sim 1\%$ at $k=0.4\kMpc$, and $\sim 5\%$ at $k_{\rm bar}=1.25\kMpc$. The Steel sample saturates at slightly worse constraining power than the Gold sample with pessimistic redshift assumptions. 

Thus, in order to improve over a sample with the number density of Steel, redshift calibration and baryon modeling must be improved beyond the assumptions used here. The \twotimestwo\ constraints do not change as a function of these parameters as they only govern impacts on the matter power spectrum. The significant improvements available in \threetimestwo\ analyses illustrate that uncertainties on these effects, as well as photometric redshift uncertainties, are likely to be the dominant source of error in future \threetimestwo\ analyses. Our model for the impact of baryons breaks down for $k>1\kMpc$, but fixing the impact of baryons to zero and relaxing our scale cuts to $\Lambda = 4\, \kMpc$ approximately preserves the relative constraining power between the Steel and the two Gold scenarios, indicating that Steel remains relatively competitive with Gold even in a future where baryonic processes and feedback are understood perfectly and scale cuts can be significantly relaxed.

\section{Observational Feasibility}
\label{sec:feasibility}
\begin{figure}[h!]
    \centering
    \includegraphics[width=\linewidth]{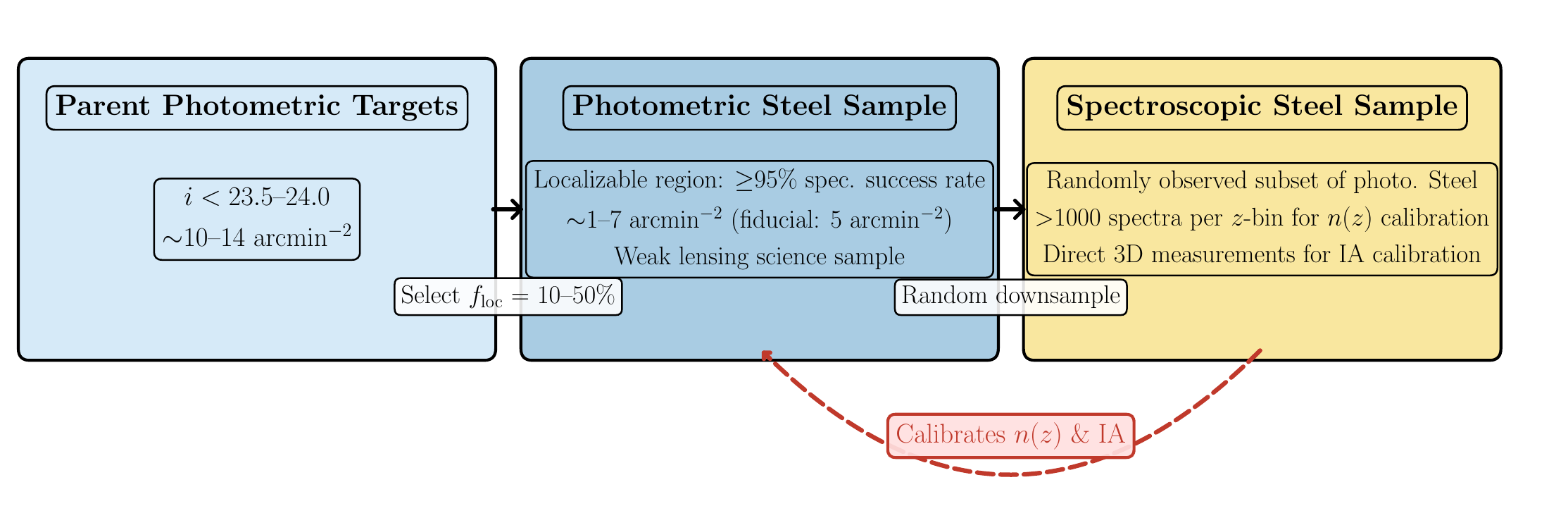}
    \caption{Schematic of the Steel sample selection strategy. The parent photometric targets are an $i$-band magnitude-limited sample. The photometric Steel sample is a localizable subset (fraction $f_{\rm loc}$) defined by a region in color-magnitude space where spectroscopic observations achieve a high success rate ($\geq 95\%$) and are representative; this is the weak lensing science sample. The spectroscopic Steel sample is a random downsample of the photometric Steel sample whose spectra calibrate the redshift distribution, $n(z)$, and IA amplitude of the photometric sample. }
    \label{fig:steel_diagram}
\end{figure}

We now perform an initial assessment of the feasibility of spectroscopically calibrating a Steel sample with a density of 5 arcmin$^{-2}$. We assess viable parent samples and the tradeoffs with magnitude limit, as well as target requirements for a spectroscopic observing program to identifying and calibrate a Steel selection from the parent sample. Figure~\ref{fig:steel_diagram} illustrates the relationship between the three key samples discussed in this section, following the sample selection approach of \cite{trudircal}. A magnitude limited sample forms the parent photometric targets, from which the photometric Steel sample is selected as the fraction, $f_{\rm loc}$, for which spectroscopic success rates of $f_{\rm spec}>95\%$ can be obtained in a given amount of effective spectroscopic observing time, $t_{\rm eff, max}$. The spectroscopic Steel sample is then a down-sampling of the photometric Steel sample that is directly observed with a spectroscopic instrument.  

We will investigate the optimal choice of $t_{\rm eff, max}$ in ref.~\cite{steel_obs}, but it is approximately a few hours of DESI effective exposure time. Our goal of $f_{\rm spec}>95\%$ spectroscopic success rates is set by a highly conservative assumption that the fraction of the sample that cannot be calibrated in $t_{\rm eff}$ will be calibratable either through deeper dedicated spectroscopy or indirect means (e.g.\ clustering redshifts) to $\sigma(\Delta z)=0.1$, thus still allowing us to achieve our requirement of $\sigma(\Delta z)=0.005$. As discussed below, the fact that we intend to use brighter sources means that follow-up to calibrate the $1-f_{\rm spec}$ fraction of sources will be significantly less expensive than doing the same for a Gold-like sample.

\subsection{Selecting a parent sample}\label{sec:parentsample}
\begin{figure}[h]
\begin{subfigure}[t]{0.5\textwidth}
    \centering
    \includegraphics[height=5.7cm]{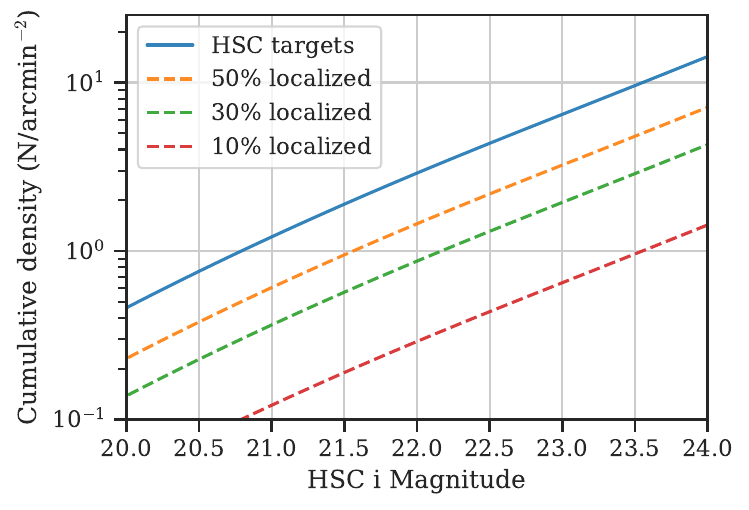}
\end{subfigure}
\begin{subfigure}{0.5\textwidth}
    \centering
    \includegraphics[height=5.7cm]{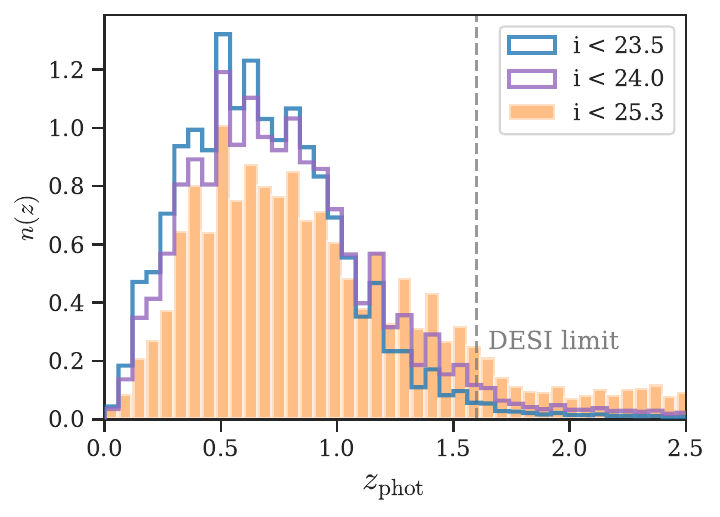}
\end{subfigure}
\caption{\textit{Left:} Cumulative density of HSC targets brighter than a given $i$-magnitude limit (solid), and the density of samples with different localization fractions (dashed).
\textit{Right:} Normalized photometric redshift distribution $n(z)$ for HSC targets with different $i$ magnitude limits. A Gold selection $i<25.3$ (filled) has roughly $16\%$ with $z_{\rm phot}>1.6$, where DESI has difficulty securing redshifts. Steel parent samples (lines) with brighter magnitude limits of $i<23.5$ (blue) or $i<24$ (purple) have significantly lower fractions at $\sim$2.5\% and 5.5\%, respectively.
}
\label{fig:dens_nz}
\end{figure}

For our photometric sample, we use HSC PDR3 Wide data \cite{Aihara_2022}, which achieves a $5\sigma$ PSF depth of $\sim26.2$ in $i$, and so serves as an excellent precursor to LSST, which is anticipated to have a corresponding 10 year $i$-band depth of $26.4$ \cite{Mandelbaum2017}. We select all HSC Wide PDR3 galaxies that pass $i$-band photometry quality cuts and are extended\footnote{We use \texttt{i\_extendedness\_value}$=1$. This removes stars from the sample but may also remove more true galaxies than is desirable, so an observational program may use a more refined star-galaxy separation scheme.}. The solid curve in the left plot of Fig.~\ref{fig:dens_nz} shows the cumulative number of targets with a given $i$-band magnitude limit. The Steel sample seeks to define a volume in color-magnitude space that can be spectroscopically calibrated with very high confidence, e.g.\ having a $\ge 95\%$ spectroscopic success rate. The ability to localize such a volume will depend on the specifics of the photometric sample, details of the spectroscopic observing program, and the number of features available (photometric bands, galaxy features, etc.). We will primarily consider using a DESI-like instrument for our spectroscopic observations, though of course the philosophy and methodological approach for defining a Steel sample is quite general.

From the ``parent" photometric targets that fall within the limited target selection criteria noted above, only a subset will actually be observed spectroscopically and our goal is to identify a subspace wherein those spectroscopic observations are representative, which we designate the photometric ``Steel" sample. We call the fraction of the parent sample that populates this calibratable photometric Steel sample the \textit{localizable fraction}; this will almost always be less than the total redshift success rate for the observed targets, as some successful redshifts will lie in regions with low local success rate and thus high likelihood of not being representative.

We show target densities for three different localization fractions as dashed lines in Fig.~\ref{fig:dens_nz}. These represent rough densities of different potential photometric Steel samples under different assumptions for the localization fraction in order to explore the space. More detailed estimates will be presented in a forthcoming paper \cite{steel_obs}.
For orientation, assuming a pessimistic localization fraction of $10\%$ for $i<23.5$ would result in a photometric Steel sample of $\bar{n}\simeq 1\,\mathrm{arcmin}^{-2}$.  At the other end a $50\%$ localization of an $i<24$ sample yields $\bar{n}\simeq 7\,\mathrm{arcmin}^{-2}$.

One generic trade-off is that, for the same investment of spectroscopic observing time, a fainter sample can be selected with more parent targets but at the cost of a lower localization fraction. For example, we find that increasing the upper limit from $i<23.5 \rightarrow i < 24$ adds $\sim50\%$ to the parent sample density (Fig.~\ref{fig:dens_nz}, left). However, in the background limited regime and for the same observing time, the population of galaxies at $i = 24$ will have roughly $10^{0.4 (24-23.5)} \approx 0.63\times$ the SNR of $i = 23.5$ galaxies, assuming similar galaxy types.  Whether this results in a net gain in Steel sample density depends strongly on $df_{\rm success}/di|_{t_{\rm obs}}$.

A related tradeoff specific to DESI is that a higher fraction of galaxies in a fainter parent sample will have $z>1.6$, where the [O{\sc ii}] doublet has redshifted out of the wavelength range of the DESI spectrographs, rendering it very hard to obtain a secure redshift. The relative size of this population has a strong influence on the fraction of the parent sample that is localizable. The right-hand plot of Fig.~\ref{fig:dens_nz} shows that while these objects account for $\simeq 16\%$ of a Gold-like sample, Steel parent samples with $i < 24$ (purple) or $i<23.5$ (blue) shrink this dramatically to 5.5\% and 2.5\%, respectively.  We use HSC \texttt{DEMP} photometric redshifts, though the conclusions do not change significantly with other HSC photo-$z$ estimators. 

Finally, even if such outliers are controlled to a low level and the $n(z)$ of each redshift bin can be accurately characterized spectroscopically, fainter samples have larger \photoz{} error, such that the $n(z)$ will be necessarily wider. Narrower source bins are preferable, as discussed previously. We therefore investigate how \photoz{} error depends on the magnitude limit of our parent sample.

\begin{figure}
    \centering
    \includegraphics[width=\linewidth]{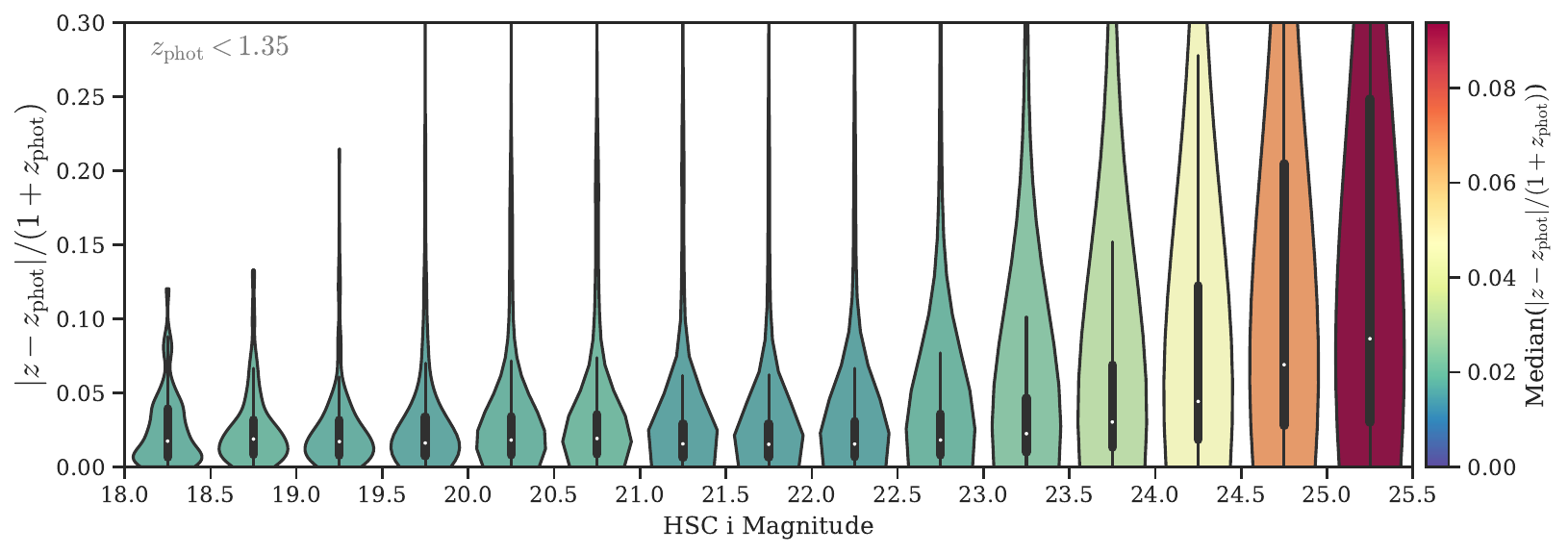}
    \caption{Violin plot of photo-$z$ errors, estimated as HSC photo-$z$ minus multi-band photo-$z$ in COSMOS in bins of HSC $i$ magnitude for galaxies with $z_{\rm photo}<1.35$. Points indicate the median and the thick inner boxes the 25-75 percentile range. 
    Distributions are colored by the median photo-$z$ error, smoothed with a $0.3\sigma$ bandwidth KDE and don't extrapolate past the raw data. Photo$z$ error increases very quickly for $i\gtrsim24$, making compact $n(z)$ selections difficult.}
    \label{fig:photo_z_error}
\end{figure}

As a proxy for deep spectroscopy, we use the many-band COSMOS2025 compilation\footnote{\url{https://cosmos2025.iap.fr/catalog.html}} \cite{Shuntov_2025} which combines data from HST, JWST, and ground based imaging in the COSMOS field. We select \lephare{} classified galaxies with no photometry warning flags and that are outside the HSC star mask and have a model $\chi^2_\lephare{}<200$.
Using a $<1^{\prime\prime}$ match radius, we cross-match these data with HSC wide galaxies with good photometry and $z_{\rm phot}<1.35$ (roughly corresponding to galaxies in all but the highest Gold redshift bin) and plot the distribution of the resulting $\sim43\,000$ objects' photo-$z$ errors $|z-z_{\rm phot}|/(1+z_{\rm phot})$ in bins of $i$, using the COSMOS 30-band redshift as truth, $z  = \texttt{zfinal}$, and $z_{\rm phot} =  \texttt{demp\_photoz\_best}$ (Fig.~\ref{fig:photo_z_error}). Each ``violin" in the plot shows the (KDE-smoothed) distribution of \photoz{} errors in that bin, with the internal box-plots showing the median (point) and inter-quartile range (IQR; 25-75th percentile; thick bars). 
Thin bars extend to points that fall within another $1.5\times$IQR past each quartile. Violins are colored according to their median \photoz{} error. Above $i>24$, the \photoz{} error begins to increase rapidly. This figure also shows that the median absolute deviation of HSC photometric redshifts is about 0.05 at $i=24$ and rapidly degrades after this, justifying the widths of the core distributions assumed for Steel shown in Table~\ref{tab:source_nz_priors} used in our forecasts.
The interplay of parent selection, redshift success rate, redshift bin width, and localizable fraction motivates future investigation. 

\subsection{Requirements on redshift survey to calibrate $n(z)$}

We estimate the number of spectra required to calibrate the Steel $n(z)$ by drawing $N_{\rm sample}$ independent samples from the true $n(z)$ of each redshift bin, and fitting\footnote{We bin the observations in $\sim N_{\rm sample}/10$ equal spaced bins and assume a variance for the counts in each bin $i$ of $\sigma^2(N_i)=N_i + 3$.} the resultant empirical $n(z)_{\rm sample}$ to Eq.~\ref{eq:nz_uncertainty} to estimate the best-fit values of {$f_{\rm out}$, $\Delta z_{\rm core},\ \sigma_{z, {\rm core}},\  \Delta z_{\rm out},\ f_{\rm out}$}, corresponding to the theoretical model used for forecasts in Sec.~\ref{sec:constraints}. This model follows that of \cite{Zhang2026}, and note that while it incorporates greater model uncertainty than a basic shift + stretch model, it still assumes that $p_{\rm core}$ and $p_{\rm out}$ are known \textit{a priori}, and so the requirements derived here are likely optimistic in this sense. We repeat this $N_{\rm trials}=100$ times and compute the root-mean-square-error (RMSE\footnote{We prefer the RMSE as a quality statistic over $\sigma(\hat{\theta})$ as it incorporates any potential biases  in the estimator.}) of the parameter estimates $\hat\theta$ over all trials:
\begin{equation} \label{eq:rmse}
    \rm{RMSE}(\theta) = \sqrt{\sum_{i=1}^{N_{\rm trials}} \left(\hat{\theta}_i - \theta_{\rm true}\right)^2},
\end{equation}
where $\theta_{\rm true}$ corresponds to the true input parameter values.

\begin{figure}
    \centering
    \includegraphics[width=\linewidth]{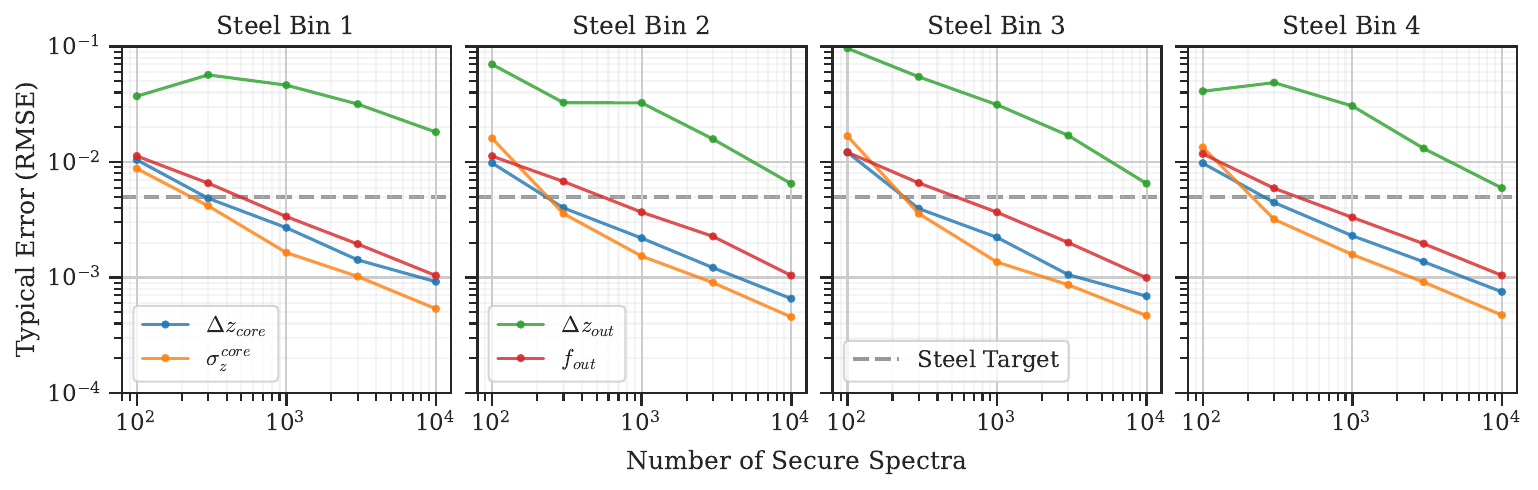}
    \caption{Estimated uncertainty on $n(z)$ nuisance parameters for the Steel sample, as a function of the number of representative spectra. RMSE is computed using Eq.~\ref{eq:rmse}, with $N_{\rm trials} = [500, 500, 300, 100, 100]$ for $N_{\rm sample} =  [100, 300, 1000, 3000, 10000]$. The dashed line gives the uncertainty assumed for our fiducial Steel constraints in Sec.~\ref{sec:constraints}.
    }
    \label{fig:steel_nz_rmse}
\end{figure}

Fig.~\ref{fig:steel_nz_rmse} shows the expected error on the parameters as a function of $N_{\rm sample}$. All bins show similar behavior, with $\sim1000$ spectra per bin comfortably achieving error rates below the fiducial $0.05$ target (dashed line, cf. Tab.~\ref{tab:params}). This sets a rough target for the total number of observed objects to be $N_{\rm obs} = 1000 \times N_{\rm bins}/ f_{\rm loc}$, or $40\,000$ for $f_{\rm loc} = 0.1$ and four redshift bins. To mitigate sample variance, these observations should be taken across several widely separated fields.  Small amounts of residual sample variance can be mitigated through standard reweighting approaches to account for chance differences between the observed fields and the full footprint, similar to existing SOMPZ approaches though with considerably reduced biases in representivity.
As an example, using DESI to observe $4000$ objects per field over 10 separated fields would provide $40\,000$ objects that are reasonably representative and have small sample variance.

We also note that there is very likely a tradeoff between $f_{\rm loc}$ and selection complexity, with higher values of $f_{\rm loc}$ for a fixed faint-end threshold likely requiring more complex selection criteria. More complex selections can be more challenging to model and replicate in simulations, or to translate to other photometric bands if using the Steel sample for cross-survey calibration purposes (a legacy use case). 

Furthermore, the more complex the Steel selection, the harder it will be to identify and mitigate variation in the selection function across the footprint due to variations in photometric calibration, seeing, PSF characterization, etc., as the dependence on theoretical templates used to identify such patterns becomes more nonlinear. This variation causes galaxies to systematically scatter in and out of the selection, directly modulating the signal for galaxy clustering and requiring significant effort to mitigate \cite{Huterer_2013, Weaverdyck:2020mff, weaverdyck2026darkenergysurveyyear}. However, fluctuations in number density will only modulate the noise of the shear measurement and so have far smaller impact, provided the variation in selection does not also couple to shape measurements. Spatial variation in redshift distributions of source galaxies do contribute at 2nd order to the cosmic shear signal, so these must be controlled to some extent \cite{Myles2021,Baleato_Lizancos_2023}.

\section{Outlook}
\label{sec:outlook}

Galaxy weak lensing is a unique probe of the low redshift matter distribution and expansion history. A number of Stage IV surveys, including LSST, Euclid and the Roman Space Telescope, have been designed to measure the shapes of very faint, dense samples of galaxies in order to reduce statistical errors on galaxy lensing measurements. This strategy is predicated on three key assumptions, namely that:
\begin{enumerate}
    \item there is significant cosmological information content in the small scale lensing modes that is not significantly degraded by the impact of baryonic physics,
    \item the redshift distributions of faint galaxies can be calibrated to sufficient accuracy so as to not bias weak lensing analyses using these galaxies,
    \item the redshift and scale dependence of the galaxy intrinsic alignment signal will not present a significant source of systematic error.
\end{enumerate}
A number of recent analyses cast significant doubt on the validity of all three of these assumptions (see e.g.\  \cite{Hadzhiyska2025,Siegel2026b,Martinelli2021,Robertson2026} for baryons, \cite{Li2023,Dalal2023,Blanco2026,trudircal} for redshift calibration and \cite{Chen2024b,Samuroff2024,Blot2025} for IAs). In this work we show that in the presence of current levels of uncertainty on baryonic physics, the information content of weak lensing analyses saturates on quasi-linear scales. This allows the use of source galaxy samples that are significantly less dense, e.g.\ with number densities of $5\rm \, arcmin^{-2}$, without sacrificing constraining power. The near optimality of such relatively sparse source galaxy samples opens the possibility to directly calibrate the redshift distributions and intrinsic alignment contamination of such a sample using a spectroscopic instrument like DESI. Such a sample would be steeled to most major weak lensing systematics, including redshift, intrinsic alignment and shear calibration as well as the impact of baryons. As such, we refer to this sample as the Steel sample. This approach follows the spirit of the \textit{true} direct calibration  (\trudircal{}) approach proposed in a parallel work \cite{trudircal} for lens samples, wherein direct spectroscopic observations of a photometric parent enables the sculpting of a final photometric sample that is well understood, with minimal contamination and tightly calibrated $n(z)$. 

In Section~\ref{sec:constraints} we use flexible models for intrinsic alignment, baryon, galaxy bias and redshift distribution uncertainties, to forecast the constraining power of combined DESI-LSST \threetimestwo\ analyses, showing that a Steel sample with a number density of $5\rm\, arcmin^{-2}$ whose redshift distributions are calibrated to $\sigma(\Delta z)=0.005$ is more constraining than the significantly denser ``Gold'' sample from LSST if this sample's redshift distributions are only calibrated to $\sigma(\Delta z)=0.01(1+z)$. This latter number may even be a slightly optimistic characterization of the current state of affairs, as it is comparable to, but slightly better than, the redshift calibrations obtained for much brighter galaxy samples in Stage-III surveys. 

This finding is relatively insensitive to the level of flexibility in the intrinsic alignment sector of our model. In particular, in Section~\ref{sec:ia_forecasts} we show that \twotimestwo\ analyses with DESI lenses can largely self-calibrate the IA contributions to our model, independent of the complexity of the redshift evolution model assumed. Because of this, tightening the priors on IA amplitudes, e.g.\  by informing them with direct IA measurements, does not add a significant amount of constraining power to the Steel sample. On the other hand, Gold constraints can be improved by $30\%$ by tightening IA priors. As the Steel sample is largely comprised of the brightest subset of Gold galaxies, direct IA constraints from the Steel sample along with assumptions about the monotonicity of IA amplitude with luminosity and color, could improve constraints using the Gold sample if its redshift distribution can be calibrated sufficiently well.

While the constraining power of \threetimestwo\ analyses is not highly sensitive to the flexibility of our IA model, we show that overly simplistic models for IA redshift evolution and scale dependence  e.g.\ assuming a single constant IA amplitude per source bin, and IA scale dependence, e.g.\ the NLA model, can lead to poor model fits and potentially significantly bias $S_8$ and $A_s$ constraints. Notably, the Steel sample is less susceptible to mis-modeling of IA redshift evolution than the Gold sample considered in this work, due to the significantly narrower redshift distributions of the Steel sample. We also show that simple extensions of these oft-used models that come at negligible cost in terms of constraining power can significantly mitigate these potential biases. In the case of IA redshift evolution, we recommend a spline redshift evolution per source bin with nodes spaced at intervals in redshift of $\Delta z=0.5$. For scale dependence, the local Lagrangian bias assumption provides goodness of fits that are very comparable to the full second order IA expansion.

In Section~\ref{sec:z_forecasts} we investigate how the $S_8$ constraining power of \threetimestwo\ analyses depends on the four degrees of freedom that we allow in each source redshift distribution. We find that the mean of the core redshift distribution is the most important parameter, in agreement with previous work, and that constraining power largely saturates at $\sigma(\Delta z_{\rm core}^i)=0.01$, although minor increases in constraining power are available by reducing this even further. Of nearly equal importance are uncertainties on the width of the core redshift distributions, parameterized by $\sigma_{z, \rm core}^i$. $S_8$ constraining power also saturates around $\sigma(\sigma_{z, \rm core}^i)=0.01$. The Gold sample is significantly more sensitive to this parameter than the Steel sample because it has many more overlapping lens-source bin pairs, and whose level of overlap strongly depends on $\sigma(\sigma_{z, \rm core}^i)$. This overlap is a key determinant in the constraining power with \twotimestwo\ analyses, and is one of the benefits of the narrow redshift bins achievable with Steel.

In our fiducial modeling scenario, we find that the constraining power of DESI-LSST \threetimestwo\ analyses saturate at a density of $\sim 5\rm\, arcmin^{-2}$ even if the redshift distributions of denser samples can be calibrated at the $\sigma(\Delta z)=0.005$ level. The most significant limitation to the constraining power of analyses using denser source samples is our lack of understanding of baryonic feedback. In a futuristic scenario, where baryons are understood at the $0.5\%$ level at $k=0.4\kMpc$, a factor of 10 better than our fiducial assumption, the constraining power of \threetimestwo\ analyses continues to improve at least until the largest densities investigated in this work ($30\rm \, arcmin^{-2}$). Beyond improvements in redshift calibration, a better understanding of galaxy formation physics and its impact on the matter distribution will be the most important requirement to make full use of upcoming Stage IV weak lensing measurements. The spectroscopic nature of the Steel sample also opens a potential door to calibrate baryonic effects using the kinetic Sunyaev-Zel'dovich effect over broader ranges of redshift and halo mass than allowed by existing samples \cite{Schaan2020,Guachalla2025,Hadzhiyska2025}. We plan on investigating this topic in the future.

We have also performed a preliminary investigation into the feasibility of calibrating a Steel sample with a number density of $5\, \rm arcmin^{-2}$ using DESI. Doing so will require us to select a sample of this density that has a spectroscopic success rate of $>95\%$ using only LSST photometry. This latter constraint is imposed so that calibrations derived in the overlap area between DESI and LSST can be applied to the full LSST footprint. Ideally, this selection will be done directly on the shear catalogs themselves. Using HSC data, we show that we can achieve our goal $5\, \rm arcmin^{-2}$ density if such a selection can localize $50\%$ of all $i<23.5$ galaxies. We also provide evidence that such a selection would have a total redshift distribution that is comparable to what we have assumed in our forecasts, and that redshift point estimates for this sample will be good enough to produce redshift bins as narrow as we have used in this work. In upcoming work we will investigate this with DESI data \cite{steel_obs}, including optimizations of this selection such as narrower source bins, higher weight toward higher redshifts, etc. 

Finally, the ultimate application of the Steel sample will be to a combined weak lensing and three dimensional clustering analysis, analogous to those performed in Ref.~\cite{Chen22,Maus25}, replacing CMB lensing with galaxy lensing data. This future direction is one of the main motivations to use spectroscopically calibrated lenses, along with the fact that the analyses forecast here are largely not shot-noise limited. Given it's high redshift nature, the Steel sample may also form a useful compliment to existing spectroscopic samples as a lens sample for CMB lensing cross-correlations. Future work will explore these avenues as additional directions for the Steel sample program.

\section*{Acknowledgments}
We thank J.~Siegel for sharing his IA measurements in electronic form.
MW thanks Nick Kokron for valuable conversations regarding IAs. NW is supported by the Chamberlain Postdoctoral Fellowship at LBNL.
MW was supported by the DOE. Support for this work was provided by NASA through
the NASA Hubble Fellowship grant HST-HF2-51572.001
awarded by the Space Telescope Science Institute, which
is operated by the Association of Universities for Research in Astronomy, Inc., for NASA, under contract
NAS5-26555. 
This research was supported in part by grant NSF PHY-2309135 to the Kavli Institute for Theoretical Physics (KITP).
This research used resources of the National Energy Research Scientific Computing Center (NERSC), a Department of Energy User Facility.

\appendix

\section{Intrinsic alignment variations}
\label{app:IA_variations}

\subsection{Scale dependence}
\label{app:ia_scale_dependence}

\begin{figure}
\resizebox{\columnwidth}{!}{\includegraphics{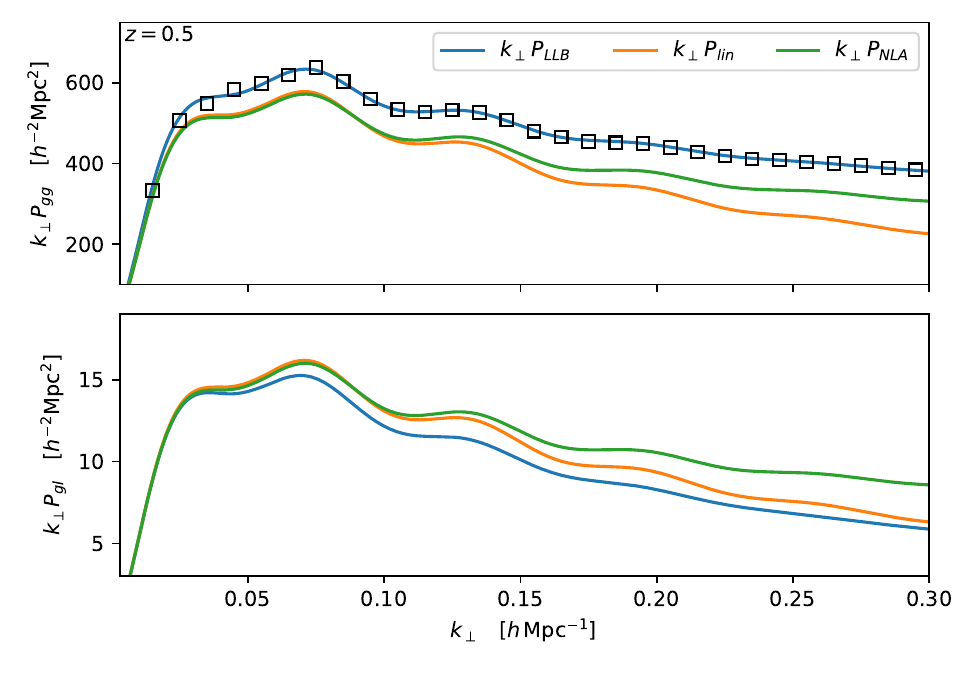}}
    \caption{A comparison of different models for IA and halo clustering.  The upper panel shows the (real-space) halo-halo auto-spectrum.  The open squares are measurements for a sample of $10^{12}\,h^{-1}M_\odot$ halos from N-body simulations at $z=0.5$, the blue line shows a fit using Lagrangian perturbation theory while the orange and green lines are $b^2P_{\rm lin}$ and $b^2P_{\rm non-lin}$ with the same value of the large-scale bias.  The lower panel shows the density-shape power spectrum for the same three models as the top panel, with the bias parameters above and the shape parameters fit to the shape-shape spectra of $10^{12}\,h^{-1}M_\odot$ halos as in Ref.~\cite{Chen2024a} (with the fit shown in Figs.~6 \& 7 of that reference, and compared to ``linear shape bias'' therein).}
    \label{fig:w_integrand}
\end{figure}

While we do not have any direct measurements (or useful upper limits) for galaxies like the Gold sample, for the case of halo shapes (where high signal-to-noise ratio measurements of alignments in simulations are possible) dropping the higher-order terms has been shown to lead to biased inferences \cite{Kurita2021,Akitsu2023,Bakx2023,Chen2024a}.  It appears reasonable to argue additional physical processes could cause galaxy shapes to be less correlated than halos, but it is unclear why the scale-dependence would be expected to be simpler or the shapes more correlated\footnote{The NLA assumes that the galaxy shape field is 100\% correlated with the present-day dark matter density field to arbitrarily small scales.  Any contributions to galaxy shapes that are not 100\% correlated with the matter field would lead to decorrelation, violating this assumption.} with the larger scale, non-linear, matter field.  Figure \ref{fig:w_integrand} compares $P_{gg}$ and $P_{gI}$ for a perturbative model with local Lagrangian bias (that has the same number of free parameters as TATT), a model using linear theory and constant bias and the NLA which assumes non-linear matter clustering but linear bias and shape response.
When the clustering is projected over a line-of-sight distance large compared to the scales being probed the conversion from $k_\perp$ to $r_\perp$ is
\begin{equation}
    w_{gg}(r_\perp) = \int\frac{k_\perp\,dk_\perp}{2\pi}\ P_{gg}(k_\perp) J_0(k_\perp r_\perp)
    \quad , \quad
    w_{g+}(r_\perp) = -\int\frac{k_\perp\,dk_\perp}{2\pi}\ P_{gI}(k_\perp) J_2(k_\perp r_\perp) .
\end{equation}
For reference $J_0$ peaks at 0 and has its first zero crossing at $k_\perp r_\perp\approx 2.4$, while $J_2$ peaks at $\approx 3.5$ and has its first zero crossing at $k_\perp r_\perp\approx 5.1$.  We see that, on scales frequently used to fit IAs, using linear shape response the differences between the models are non-negligible.

The open squares in the upper panel of Fig.~\ref{fig:w_integrand} are real-space power spectrum measurements for a sample of $10^{12}\,h^{-1}M_\odot$ halos from N-body simulations at $z=0.5$.  The lines compare the theoretical models, and show the very well-known result that scale-independent bias cannot provide a decent fit to halo clustering beyond the linear regime \cite{Barnes1985,White1987}.  The lower panel shows $P_{gI}$ for the same three models as the top panel, with the bias parameters fit to the density-density power spectrum and the shape parameters fit to the shape-shape spectra of $10^{12}\,h^{-1}M_\odot$ halos as in Ref.~\cite{Chen2024a}.  For reasons of space, we do not reproduce the shape-shape spectra in Fig.~\ref{fig:w_integrand}.  They can be found Figs.~6 \& 7 of ref.~\cite{Chen2024a}, which compare perturbation theory fits to NLA (which assumes a ``scale-independent shape bias'') and tell a qualitatively similar story to the top panel of Fig.~\ref{fig:w_integrand}.

Figure \ref{fig:w_integrand} shows that the more complete models have a different scaling than the NLA, though that difference is modest on the scales we plot (a finding also seen in hydrodynamic simulations, see ref.~\cite{Herle2026} for a recent example).  For this reason, when the IA contribution to the total signal is small the modest error introduced by NLA is frequently neglected (though see ref.~\cite{Samuroff2024} for possible implications).  An alternative approach would be to use a more complex model (e.g.\ the TATT model or the LLB model shown in Fig.~\ref{fig:w_integrand}) but include informative priors on the additional parameters.  This helps to limit potential projection effects, while providing a straightforward way for the data to indicate tension with the model assumptions.  We leave further exploration of this to future work.

Of course the situation is different for measurements where the IA is not a small contribution to the signal.  This includes `direct' IA measurements, where they are the majority of the signal.  In this situation the use of the more complete models would be expected to lead to more accurate conclusions about the relationship between the galaxy shear field and the underlying matter distribution.  The necessity of introducing priors on the higher order bias parameters would be dictated by the quality of the data being fit.  Another example is when the IAs are being ``self calibrated'' by the overlapping source and lens bins in the \twotimestwo\ analyses, such as in the main text.  Since the IA contribution is non-negligible for those bin pairs, an accurate modeling of the IA signal is important.  As we have seen above, in the Stage IV era we will have enough data to fix the more complex scale- and redshift-dependent models and their use can significantly reduce modeling-induced biases.



\subsection{Redshift evolution}
\label{app:ia_z_dependence}

\begin{figure}
    \centering
    \includegraphics[width=\columnwidth]{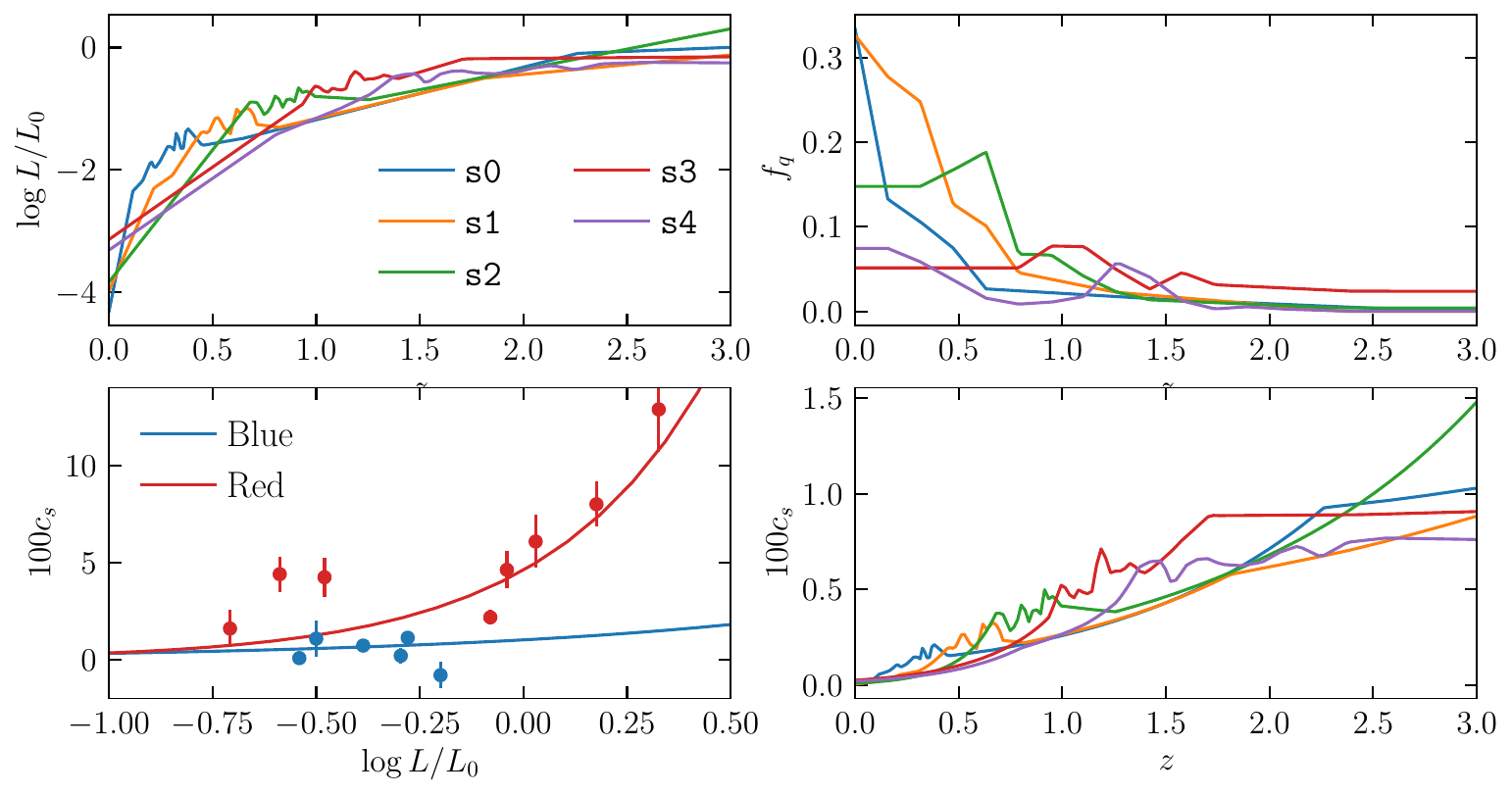}
    \caption{(\textit{Top-left}) Luminosity evolution as a function of redshift for Gold-like galaxies in five source bins (colors) as measured from COSMOS2025. (\textit{Top-right}) Quenched fraction evolution as a function of redshift for Gold-like galaxies as measured from COSMOS2025. (\textit{Bottom-left}) Assumed relationship between $c_s$ and $L$ for red and blue galaxies (lines), compared with recent direct measurements (points). (\textit{Bottom-right}) Predicted $c_s$ evolution as a function of redshift from Equation~\ref{eq:ia_z_model}, assuming the inputs shown in the other panels. See details in App.~\ref{app:IA_variations}.}
    \label{fig:ia_contamination}
\end{figure}

As was the case for all of the Stage III lensing surveys, the $n(z)$ for the proposed Gold sample galaxies of LSST span a non-trivial range in redshift both in the core and for the outliers (see Fig.~\ref{fig:nz}).  These galaxies are in general selected by a complex series of color and magnitude cuts, and thus we fully expect the mix of galaxy types and their properties to evolve across the sample.  Such an evolution modifies the signals of interest in a way that depends upon the kernels that enter into Eq.~(\ref{eq:limber}).

Of particular interest to us is the degree to which any possible IA signal evolves with redshift for a given source bin.  Unfortunately, there are no direct IA measurements for the Gold sample galaxies, only upper limits (see e.g.\ Fig.~9 of \cite{Siegel2026a}).  Further, even for the galaxies where we have upper limits, those limits are loose enough that IAs could still be a major issue for shear auto-correlations at the level of precision of Stage IV surveys.

While we cannot show that IA evolution across the sample is negligible, we can make some rough estimates to attempt to bound the possible size of the effect.  An expected evolution is of the galaxy luminosity at fixed flux: the luminosity limit associated with a specific sample should scale with $z$ (approximately) as $d_L^2$, where $d_L$ is the luminosity distance.  Even $\Delta z=0.5$ can lead to $\mathcal{O}(1)$ variations in luminosity for the samples shown in Fig.~\ref{fig:nz}.  Earlier observations have indicated that for the brighter, redder galaxies for which IA can be detected the amplitude of the alignment signal is approximately linear in luminosity \cite{Fortuna2021,Samuroff2023}.  If such a dependence, albeit at much reduced amplitude, also held for the fainter, bluer galaxies this would require us to include per-sample IA evolution in our forecasts. Even a significantly weaker trend with luminosity would result in loose enough priors that evolution would need to be incorporated. 

In order to quantify the importance of modeling this evolution, we have constructed a data-driven IA redshift evolution scenario that assumes the IA amplitude depends only on luminosity and whether a galaxy is star-forming (blue) or not. We assume a power-law functional form for the dependence of IA amplitude on luminosity:
\begin{equation}
    \tilde{c}_s(L) = c_{s,0}\left(\frac{L}{L_0}\right)^{\beta}\,
\end{equation}
where we set $c_{s,0}^{\rm red}=0.046$, $\beta^{\rm red}=1.13$, and $c_{s,0}^{\rm blue}=0.010$, $\beta^{\rm blue}=0.5$ from Table 4 of \cite{Joachimi2012} and the $z>0.5$ blue sample from \cite{Paviot2026} respectively, converting between $A_1$ and $c_s$ appropriately. $L_0$ corresponds to an absolute magnitude of $M_r=-22$. We then use HSC photometric redshifts (\texttt{dnnz\_photoz\_best}) to bin the cross-matched COSMOS2025 galaxies described in Sec.~\ref{sec:parentsample} into Gold and Steel-like redshift bins, applying an apparent magnitude threshold of $m_i<25.3$ and $m_i<23.5$ respectively. In each bin, we measure the luminosity of each sample as a function of COSMOS 30-band photometric redshift, $L^{i}(z)$, as well as the quenched fraction of galaxies, $f_{q}^i(z)$. The quenched fraction is defined as the fraction of galaxies with specific star formation rate $\rm SSFR<10^{-11}\rm \, yr^{-1}$. We then make predictions for the redshift evolution of $c_s$ for each source bin as:
\begin{equation}
    c_s^{i}(z) = f_{q}^i(z)c_{s}^{\rm red}\left(L^{i}(z)\right) + (1-f_{q}^i(z))c_{s}^{\rm blue}\left(L^{i}(z)\right)
    \label{eq:ia_z_model}
\end{equation}

Figure~\ref{fig:ia_contamination} shows the various pieces of our model. The top-left and top-right panels shows our measured $L^{i}(z)$ and $f_q^i(z)$ relations from the COSMOS data for a Gold-like sample for each source bin. The bottom-left panel shows the assumed $\tilde{c}_s(L)$ relations, as well as a compilation of NLA amplitude measurements from direct measurements \cite{Joachimi2012,Samuroff2023,Girones2026,Siegel2026a} for red and blue galaxies. The bottom right panel shows the resulting $\tilde{c}_s^i(z)$ for our Gold-like sample. We apply an analogous procedure to construct $\tilde{c}_s^i(z)$ for a Steel-like sample. We use these for our IA model complexity tests in Section~\ref{sec:constraints}.

This procedure results in an obvious trend in $\tilde{c}_s$ with redshift for all source bins, evolving from $\tilde{c}_s\sim0.01$ to $\tilde{c}_s\sim 0$ between $z=2$ and $z=0$. At a given redshift, the source bins do exhibit differences in luminosity and quenched fraction, but because the slope of $\tilde{c}_s(L)$ is shallow for the faint luminosity range that is largely probed by weak lensing source samples, these differences in luminosity translate into relatively minor differences in $\tilde{c}_s$ at fixed redshift between the bins. The key assumptions that dictate the amount of redshift evolution seen in each bin, and the amount of variation between bins are:
\begin{itemize}
    \item $\tilde{c}_s$ depends only on luminosity and quenched fraction.
    \item $\tilde{c}_s(L)$ is a power law, monotonically decreasing to zero at faint luminosities.
    \item COSMOS2025 30-band \photoz\ estimates allow us to accurately characterize the luminosity and quenched fraction evolution of very faint galaxies.
\end{itemize}
\noindent This last point is particularly relevant, since one could imagine that if the COSMOS \photoz estimates have a significant outlier fraction for faint magnitudes, then this outlier population may have a significantly different luminosity or quenched fraction than the in-distribution galaxies, thus leading to potentially pathological IA evolution that is not seen in our current model.

\section{Direct IA measurement}
\label{app:direct_ias}
Direct measurements of intrinsic alignments by cross-correlating galaxy shapes and galaxy positions, where both shape and positions samples have been binned narrowly in redshift, are the only direct mechanism that we have for measuring the intrinsic alignment signal. Ever since the IA signal was identified as being a significant source of error in weak lensing analyses \cite{Catelan2001}, spectroscopic galaxy samples have been used to estimate the size of the contamination \cite{Mandelbaum2006,Hirata2007,Joachimi2012,Singh2015,Samuroff2023,Siegel2026a}. Recently, complete effective field theory descriptions of IAs have been formulated \cite{Vlah2020,Chen2024a} and applied to simulations \cite{Bakx2023,Vedder2026} and data \cite{Chen2024b,DeRose2025}. 

In the main text, we discuss avenues for controlling biases in cosmological constraints due to mis-modeling of the redshift- and scale-dependence of the IA signal. We conclude that for the Steel sample, the level of model complexity required in the IA sector can be sufficiently self-calibrated in \threetimestwo\ analyses, but for the Gold sample we show that significant improvements in IA constraints over the self-calibration ability of \threetimestwo\ analyses can improve $S_8$ constraining power. In particular, a direct measurement with $\sigma(\tilde{c_s})=0.002$ ($\sim\sigma(A_1)=0.25$) would lead to a $30\%$ improvement in $S_8$ constraining power for a Gold like analysis with optimistic redshift calibration. Even though direct spectroscopic calibration of the Gold sample is infeasible, measurements of IAs using the Steel sample may be able to contribute indirect constraints on the IA contributions in the Gold sample, for example by measuring the brightest subset of Gold galaxies, and assuming monotonicity of the IA signal with luminosity and/or color \cite{McCullough2024b,Bigwood2026}. Towards this end, in this appendix, we investigate how tightly we can constrain the IA amplitude of the Steel sample, taking full advantage of three-dimensional direct IA measurements which may provide significant advantages \cite{Singh2023,Lamman2025} over the projected measurements that are usually employed \cite{Joachimi2012,Singh2015,Siegel2026a}. 

\subsection{Forecasting formalism}

In order to forecast our ability to constrain intrinsic alignments using spectroscopic samples we will need a model for the clustering of IAs in redshift space. In this case the helicity basis is no longer convenient since the line of sight direction $\hat{n}$ breaks the rotational symmetry in real-space, and EFT predictions for redshift-space IAs have not been fully developed (but see e.g.\ \cite{Taruya2025}). However, general considerations of symmetry and the structure of nonlinearities lets us write an empirical model for clustering between galaxy densities and shapes as
\begin{equation}
    \delta_{g,s}(\bk) = (b_1 + f\mu^2) T_{g}(k,\mu) \delta_{\rm lin}(\bk) + \epsilon(\bk), \quad g_{s,ij}(\bk) = c_s T_{I}(k,\mu) s_{{\rm lin}, ij}(\bk) + \epsilon_{ij}(\bk).
\end{equation}
where the transfer functions are given by 
\begin{equation}
    T_{g,I}(\bk) = 1 + \sum_{n,m>0} a_{n,m}^{g,I} \left( \frac{k}{k_{\rm nl}} \right)^{2n} \mu^{2m}
\end{equation}
and $\epsilon, \epsilon_{ij}$ describe mode-coupling and stochastic contributions not directly proportional to the linear power spectrum. Here, $k_{\rm nl}$ is the nonlinear scale at a given redshift defined such that $\Delta^2(k_{\rm nl}) = 1$; below this scale, nonlinearities are expected to be perturbative additions to linear theory, as is reflected in the parametrization above when $a_{n,m} \lesssim 1.$ The redshift-space power spectra of galaxy densities and shapes ($E$- and $B$-modes) is then given by
\begin{align}
    &P^{\delta_g\delta_g}(\bk) = (b_1 + f\mu^2)^2 T_g^2(\bk) P_{\rm lin}(k) + P_{\epsilon}^{\delta_g\delta_g}(\bk) \nonumber \\
    &P^{\delta_g g^E}(\bk) = \frac{c_s}{2} (1-\mu^2) (b_1 + f\mu^2) T_g^2(\bk) P_{\rm lin}(k) + P_{\epsilon}^{\delta_g g^E}(\bk) \nonumber \\
    &P^{g^Eg^E}(\bk) = \frac{c_s^2}{4} (1-\mu^2)^2 P_{\rm lin}(k) + P_{\epsilon}^{g^Eg^E}(\bk)  \nonumber \\
    &P^{g^Bg^B}(\bk) =  P_{\epsilon}^{g^Bg^B}(\bk), \quad P^{\delta_g g^B}(\bk) = P^{g^Eg^B}(\bk) = 0.
    \label{eqn:rsd_pk}
\end{align}
Similarly to the transfer functions we can generically write the mode-coupling and stochastic contributions as
\begin{equation}
    P_{\epsilon}^{\alpha \beta}(\bk) = \sum_n \sum_{m \leq n} s_{n,m}^{\alpha \beta} \left( \frac{k}{k_{\rm nl}} \right)^{2n} \mu^{2m}
\end{equation}
where for example $s^{\delta_g\delta_g}_{0,0}$ is the galaxy shot noise. In this case symmetry limits $s^{g^Eg^E}_{0,0} = s^{g^Bg^B}_{0,0}$ (i.e.\ the shape noise) and $s^{\delta^gg^E}_{0,0} = 0$, with no further restrictions on coefficients with higher order scale dependence \cite{Chen2024a}. This implies that any improvements in measuring the linear IA amplitude from including $P^{g^Bg^B}$ come from better constraining the scale-independent shape noise, which is consistent with the observation in the EFT of IAs that more than two nonlinear operators can independently contribute to the helicity $m=1,2$  spectra in real-space. Finally, we note that the scale dependence of the mode-coupling and stochastic pieces can be in principle described as a combination of the nonlinear scale (former) and inter-galaxy separation $R_h^{-1}$---however, we use the larger former scale above to be conservative.

\begin{figure}
    \centering
    \includegraphics[width=\linewidth]{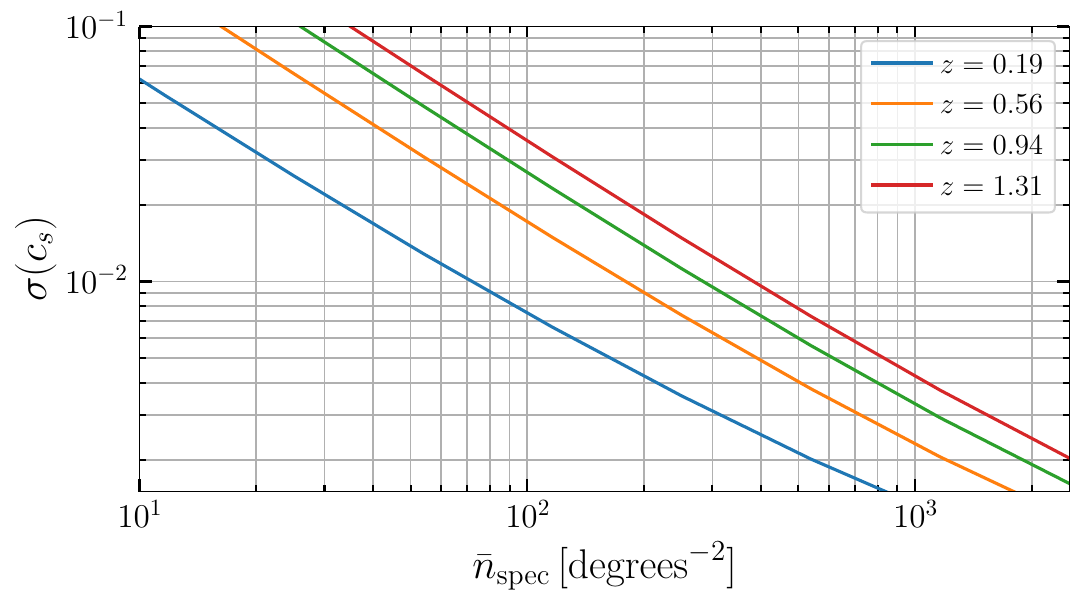}
    \caption{Forecasts of the constraining power on the linear IA amplitude, $c_s$, as a function of galaxy number density and redshift (shown by different colored lines) for a survey area of $500\, \deg^{2}$. These constraints are almost entirely shot noise dominated, and so scale like $\sigma(c_s) \propto \bar{n}_{\rm spec}^{-1}A^{-1/2}$.}
    \label{fig:cs_forecast}
\end{figure}

\subsection{Requirements on redshift survey to calibrate IAs}

We now turn to an investigation of how large a redshift survey is required to calibrate the intrinsic alignment contributions to the Steel sample. In order to forecast this, we assume that the Steel sample will be divided into four redshift bins split equally in redshift between $z=0$ and $z=1.5$, with the same angular number density in each bin. In order to observe these galaxies with DESI we will necessarily have to downsample their density by a large factor, as DESI's field-of-view is 8 deg$^2$ with approximately 4300 fibers available for extragalactic targets. 

We now wish to forecast how tightly we can constrain $c_s$ as a function of survey area and angular number density of the spectroscopic Steel sample. For this we use a field level Fisher formalism assuming redshift space galaxy density and shape fields $\delta_{g,s}(\vec{k})$ and $g_{s,ij}(\vec{k})$ rather than the 2-point function Fisher formalism used for our lensing forecasts \cite{Tegmark1997}:
\begin{equation}
    F_{ij} = \frac{\delta_{mn}^{K}}{2}\ C_{m}^{-1}\partial_{i} C_{m}\ C^{-1}_{n}\partial_{j}C_{n}\, ,
\end{equation}
\noindent where $m$ and $n$ index over bins of $k$ and $\mu$ and the covariance of a Gaussian field is given by the redshift space power spectra described in Equation~\ref{eqn:rsd_pk}, i.e.

\begin{equation}
[C_{m}]^{XY} = \frac{2\pi^2}{V_{\rm eff}\Delta \mu \Delta k\, k^2}
\begin{cases} 
      P^{\delta_{g,s}\delta_{g,s}}(\vec{k}_m)  & X=\delta_{g,s}(\vec{k}_m)\, , Y=\delta_{g,s}(\vec{k}_m) \\
      P^{\delta_{g,s}g^{E}_{s}}(\vec{k}_m)  & X=\delta_{g,s}(\vec{k}_m)\, , Y=g^{E}_{s}(\vec{k}_m) \\
      P^{g^{E}_{s}g^{E}_{s}}(\vec{k}_m)  & X=g^{E}_{s}(\vec{k}_m)\, , Y=g^{E}_{s}(\vec{k}_m) \\
      0  & X\in\{\delta_{g,s}(\vec{k}_m)\, ,g^{E}_{s}(\vec{k}_m)\}\, , Y=g^{B}_{s}(\vec{k}_m)       \\     
      P^{g^{B}_{s}g^{B}_{s}}(\vec{k}_m)  & X=g^{B}_{s}(\vec{k}_m)\, , Y=g^{B}_{s}(\vec{k}_m)      
\end{cases}
.
\end{equation}
For a given area we take $V_{\rm eff}$ to be independent of $\vec{k}$, i.e.\ we neglect window effects, other than setting the minimum wavenumber included in our Fisher matrix to $k_{\rm min}=2\pi/(\tan(\sqrt{A})\chi(z_{\rm eff}))$, where $A$ is the survey area in sr$^2$. We take $\Delta \mu=\frac13$ and $\Delta k=0.01\,\kMpc$ and $k_{\rm max}=0.3\,\kMpc$, where the latter choice is motivated by the findings in Ref.~\cite{Chen2024a,Bakx2023}. We note that we use the convention where the auto-spectra all contain noise contributions, i.e.\ $N^{\delta_g\delta_g} = 1/\bar{n}_i$ and $N^{\gamma_E^i\gamma_E^i} = \sigma_e^2/\bar{n}$, where now $\bar{n}$ is the three dimensional galaxy density which is taken to be the same for the shape and density samples.

Figure~\ref{fig:cs_forecast} shows the results of these forecasts for a 500 deg$^{2}$ survey, with the number density of each of the four spectroscopic Steel sample redshift bins on the $x$-axis and the forecast constraining power on $c_s$ on the $y$-axis. The colored lines indicate the four different redshift bins. In general, these constraints are highly shape noise dominated so they scale roughly as $A^{-1/2}\, N^{\gamma_E\gamma_E}$.

While this behavior is likely dependent on the parameterization that we have assumed for redshift uncertainties, it does provide a potentially useful target for $c_s$ constraints. Because we are in the shape noise-dominated regime, the highest redshift bin is always the most difficult to calibrate at fixed angular number density, given that it has the largest effective volume. Requiring that $\sigma(c_s)<0.02$ for this bin for a 500 deg$^2$ survey requires an angular number density of $\bar{n}=200\, \rm deg^{-2}$. For a dedicated survey, spectroscopic observing time scales as $T_{\rm obs} \propto A \bar{n}$, assuming that denser samples are not more difficult to obtain redshifts for, which should hold since even the densest samples shown in Figure~\ref{fig:cs_forecast} are still significantly less dense than 1 $\rm arcmin^{-2}$ so that density can be increased without changing the photometric Steel sample that is being downsampled to obtain spectroscopic targets. For a fixed $T_{\rm obs}$, this drives one to less area and larger densities, with the large caveat that this places significantly more importance on the accurate modeling of scales near $k_{\rm max}=0.3\kMpc$. 

\section{NLA vs.\ EFT}
\label{app:IA_conventions}

The (non-)linear alignment model, TATT and Lagrangian perturbation theory all agree on the lowest order contribution to IA, i.e.\ on the very large scale limit.  For NLA and TATT this is usually written in Eulerian space as $\gamma_{ij}^{\rm TATT} = C_1^{\rm TATT} s_{ij} + \cdots$ while for PT $M_{ij}^{\rm Lag} = c_s^{\rm Lag} s_{ij} + \cdots$ in Lagrangian space (i.e.\ the initial conditions).  On very large scales we can neglect the displacements due to evolution (which are small compared to the wavelength of the perturbation).  This means that as $k\to 0$ we have \cite{Blazek2019}
\begin{equation}
    P_{gE} = C_1^{\rm NLA, TATT} b\, P_{\rm lin} + \cdots
    \quad , \quad
    P_{EE} = \left(C_1^{\rm NLA, TATT}\right)^2 P_{\rm lin} + \cdots .
\end{equation}
while, for perturbation theory
\begin{equation}
    P_{g E}^{\rm Lag} = \frac{c_s}{2} b\, P_{\rm lin} + \cdots
    \quad , \quad
    P_{E E}^{\rm Lag} = \frac{c_s^2}{4} P_{\rm lin} + \cdots
\end{equation}
with $c_s=2\,c_1$, the factor of two difference arising due to the convention\footnote{To lowest order the shear is $[\hat{k}_i\hat{k}_j-(1/3)\delta_{ij}]\delta(\mathbf{k})$.  The E-mode of the shear can be obtained from $(1/2)[ m^{+}_i s_{ij} m^{+}_j + m^{-}_i s_{ij} m^{-}_j]$ with $m_{\pm}=(\mp\hat{x}-i\hat{y})/\sqrt{2}$.  This gives e.g.\ $\gamma=(1/2)(\hat{k}_x^2-\hat{k}_y^2)\delta(\mathbf{k})$ with the $1/2$ prefactor responsible for the difference in conventions.} used for projected shapes in the definition of the NLA model (e.g.\ eq.~14 of Ref.~\cite{Blazek2019}) compared to the three dimensional definition of the shape tensors employed in EFT models (e.g.\ eq.~3.16 of Ref.~\cite{Vlah2021}). 

Note the very well known fact that IA and cosmic shear are perfectly degenerate at linear order (i.e.\ both proportional to $P_{\rm lin}$).  It has become conventional to replace $C_1$ with
\begin{equation}
    C_1^{\rm TATT} = C_1^{\rm NLA} \simeq  -A_{\rm IA} \frac{0.01388\,\Omega_m}{D(z)}
\end{equation}
with $A_{\rm IA}$ a number of order unity.  Obviously the above gives $c_s$ as twice this.

%

\bibliographystyle{JHEP}
\bibliography{main}
\end{document}